\documentclass[         %
aps,                    
prd,                    
showpacs,               
superscriptaddress,    
nofootinbib,            
showkeys,               %
preprintnumbers,        %
floatfix]               
{revtex4}               
\usepackage{graphicx,longtable,amsmath} 
\usepackage[bbgreekl]{mathbbol} 
 
\numberwithin{equation}{section}

\begin{document} 
\title{Invariants of Collective Neutrino Oscillations} 
\author{        Y. Pehlivan} 
\email{         yamac@physics.wisc.edu} 
\affiliation{Mimar Sinan Fine Arts University, Istanbul 34349, Turkey} 
\affiliation{National Astronomical Observatory of Japan 2-21-1 
Osawa, Mitaka, Tokyo, 181-8588, Japan} 
\author{        A.~B. Balantekin} 
\email{         baha@physics.wisc.edu} 
\affiliation{Department of Physics, University of Wisconsin - Madison, Wisconsin 53706 USA } 
\author{        Toshitaka Kajino} 
\email{         kajino@nao.ac.jp} 
\affiliation{National Astronomical Observatory of Japan 2-21-1 
Osawa, Mitaka, Tokyo, 181-8588, Japan} 
\affiliation{Department of Astronomy, University of Tokyo, Tokyo 113-0033, Japan} 
\author{        Takashi Yoshida} 
\email{         tyoshida@astron.s.u-tokyo.ac.jp} 
\affiliation{Department of Astronomy, University of Tokyo, Tokyo 113-0033, Japan} 
\date{\today} 
\begin{abstract} 
 
We consider the flavor evolution of a dense neutrino gas by taking into account both vacuum oscillations and self interactions of neutrinos. We examine the system from a many-body perspective as well as from the point of view of an effective one-body description formulated in terms of the neutrino polarization vectors. We show that, in the single angle approximation, both the many-body picture and the effective one-particle picture possess several constants of motion. We write down these constants of motion explicitly in terms of the neutrino isospin operators for the many-body case and in terms of the polarization vectors for the effective one-body case. The existence of these constants of motion is a direct consequence of the fact that the collective neutrino oscillation Hamiltonian belongs to the class of Gaudin Hamiltonians. This class of Hamiltonians also includes the (reduced) BCS pairing Hamiltonian describing superconductivity. We point out the similarity between the collective neutrino oscillation Hamiltonian and the BCS pairing Hamiltonian. The constants of motion manifest the exact solvability of the system. Borrowing the well established techniques of calculating the exact BCS spectrum, we present exact eigenstates and eigenvalues of both the many-body and the effective one-particle Hamiltonians describing the collective neutrino oscillations. For the effective one-body case, we show that spectral splits of neutrinos can be understood in terms of the adiabatic evolution of some quasi-particle degrees of freedom from a high density region where they coincide with flavor eigenstates to the vacuum where they coincide with mass eigenstates. We write down the most general consistency equations which should be satisfied by the effective one-body eigenstates and show that they reduce to the spectral split consistency equations for the appropriate initial conditions.  
 
\end{abstract} 
\medskip 
\preprint{} 
\maketitle 
 
\vskip 1.3cm 
 
\section{Introduction} 
\label{Section: Introduction} 

Collective oscillations of neutrinos is the process by which neutrino-neutrino scattering contributes to the flavor evolution of  a sufficiently dense neutrino gas in a way which is somewhat similar to the well known matter-enhanced neutrino oscillations in the Sun. Such high neutrino densities are believed to be achieved in the core-collapse supernovae \cite{Qian:1994wh,Qian:1995ua,Pastor:2002we,Balantekin:2004ug,Fuller:2005ae}, in the Early Universe \cite{Kostelecky:1993yt,Kostelecky:1993ys,Abazajian:2002qx,Ho:2005vj}, and possibly in other astrophysical sites. The contribution of the neutrino-neutrino scattering to the flavor evolution differs from that of ordinary matter scattering due to the fact that in the former the scattered and the scattering particles are of the same kind, giving rise to exchange type forward scattering terms \cite{Pantaleone:1992eq,Pantaleone:1992xh}. These terms couple the flavor evolutions of neutrinos with different energies and turn the study of the system into a nonlinear many-body problem. A \emph{mean-field} type effective one-particle approximation to this problem was proposed in Refs. \cite{Pantaleone:1992eq,Pantaleone:1992xh} and has been widely adopted in the subsequent studies. However, it became evident that analytical solutions of the resulting non-linear evolution equations were needed in order to explore the full range of physical scenarios. Recent analytical studies of various special cases, such as the limit in which neutrino-neutrino scattering potential becomes dominant, revealed several situations in which neutrinos with different energies oscillate collectively \cite{Fuller:2005ae,Kostelecky:1994dt,Samuel:1995ri,Friedland:2003dv,Friedland:2003eh,Duan:2005cp,Friedland:2006ke,Duan:2006an,Hannestad:2006nj,Raffelt:2007cb}.  
 
An algebraic approach to the problem was worked out in Ref. \cite{Balantekin:2006tg} from a many-body point of view. In this approach the flavor evolution of the many-body system is formulated as an SU(2) or SU(3) coherent state path integral for two or three flavors, respectively.  The evolution operator for the entire system is calculated using both the saddle point and the operator product linearization approximations and these two methods were shown to yield the same answer. Such an approach is useful in providing a framework to look for exact solutions or systematic approximations. 
 
There is an increasingly growing literature studying the collective neutrino oscillations (see, e.g., Refs. \cite{Duan:2009cd} through \cite{Raffelt:2011yb}). There are, however, several excellent recent reviews that may serve as a starting point in exploring this literature \cite{Duan:2009cd,Duan:2010bg,Raffelt:2010zza}.  
 
The algebraic approach proposed in Ref. \cite{Balantekin:2006tg} is helpful in exploring the hidden symmetries of the system. Hamiltonian describing collective neutrino oscillations possesses an $SU(N)_f$ rotation symmetry in the neutrino flavor space \cite{Balantekin:2006tg,Duan:2008fd,Balantekin:2009dy}. Various collective modes, including spectral swappings or splittings arise from this symmetry even in the inhomogeneous  or anisotropic environments \cite{Duan:2008fd}. One expects that such a complex nonlinear system may exhibit further symmetries. Indeed, several authors noted the presence of various conserved quantities in collective neutrino oscillations \cite{Raffelt:2007cb,Duan:2007mv}.  
More recently, it was shown that collective oscillations that maintain coherence can be classified by a number of linearly-independent functions \cite{Raffelt:2011yb}, implying that scalar products of a unique linear combination of the original polarization vectors  are conserved.  
The goal of this paper is to further explore symmetries and conserved quantities associated with the collective neutrino oscillation Hamiltonian.  
 
Our study of the symmetries of the neutrino Hamiltonian is based on the observation that the neutrino-neutrino forward scattering Hamiltonian has the form of a spin-exchange interaction as was pointed out by many authors earlier. Here, the spin does not refer to the intrinsic spin of the neutrino but to the so called \emph{neutrino isospin} which is defined by introducing a multiplet of neutrino states. Interactions of this type are also encountered in many-body systems with pair coupling where the role of the spin is played by the so called \emph{quasi-spin}. Examples include the residual pairing interaction between nucleons in the nuclear shell model and the pairing of valance electrons in the BCS theory of superconductivity \cite{Bardeen:1957mv}. The fact that the pairing Hamiltonian is exactly solvable, which hints at the existence of symmetries and associated constants of motion, was pointed out as early as 1963 by Richardson \cite{Richardson1}. These constants of motion were later identified by Gaudin \cite{Gaudin1, Gaudin2} and others \cite{Cambiaggio} (for a review, see Refs. \cite{Dukelsky:2004re,Sierra:2001cx}).  
 
Here, we consider both the exact many-body Hamiltonian and the effective one-body Hamiltonian describing collective oscillations of neutrinos. We formulate the effective one-body picture in terms of the random phase approximation (RPA) method whereby the intrinsic consistency requirements are manifested as equations of motion of the neutrino polarization vectors. Our analysis includes both the vacuum oscillations and the self interactions of neutrinos. We show that, under the single angle approximation, both the exact many-body and the effective one-particle pictures possess many constants of motion which were not so far carefully studied. We express these constants of motion in terms of the neutrino isospin operators in the case of the exact many-body Hamiltonian and in terms of the neutrino polarization vectors in the case of the RPA evolution. In both cases, the constants of motion depend on the parameter which couples the flavor evolution of neutrinos with different energies manifesting the existence of associated dynamical symmetries. 
 
We give analytical expressions for the exact eigenstates and energy eigenvalues of both the many-body Hamiltonian and the effective one-particle Hamiltonian. To achieve this goal, we use the method of Bethe ansatz \cite{Bethe}. For the RPA Hamiltonian we use the method of Bogoliubov transformations to bring the system into a diagonal form in terms of noninteracting \emph{quasi-particle} states.  
We show that the quasi-particle picture is useful in offering a formal and intuitive description of the spectral splits which were reported to occur in various numerical simulations when the neutrinos adiabatically evolve from a region of high neutrino density to the vacuum. Our explanation is complementary to the one offered in Ref. \cite{Raffelt:2007cb} in terms of the neutrino polarization vectors. A preliminary account of our results was given in \cite{Pehlivan:2010zz}.  
 
The  organization of this paper is as follow: In Section \ref{Section: The Isospin Formulation of the Problem}, we briefly review the concept of neutrino isospin and write down the many-body Hamiltonian describing the collective neutrino oscillations in terms of the isospin operators. We also present the many-body constants of motion in terms of the isospin operators in this section. In Section \ref{Section: Diagonalization of the Hamiltonian}, we present the exact many-body eigenstates and eigenvalues which are found in an analytical way using the method of Bethe ansatz. In Section \ref{Section: The Random Phase Approximation}, we apply the method of RPA to bring the many-body neutrino Hamiltonian into its effective one-particle form and we briefly review how the RPA consistency requirements yield the time evolution equations of the neutrino polarization vectors. We write down the constants of motion of the RPA evolution in terms of the polarization vectors. In this section, we also consider the eigenstates and eigenvalues of the RPA Hamiltonian and write down the RPA consistency equations for these eigenstates. At the end of this section we give a brief interpretation of the spectral splits in terms of the adiabatic evolution of quasi-particles. The first four sections of this paper deal only with neutrinos in order to keep the formulas simple and emphasize the underlying physics. We include the antineutrinos in Section \ref{Section: Antineutrinos} and describe how the results of the earlier sections, including the  invariants, are generalized in this case. We conclude the paper in Section \ref{Section: Conclusions} by elaborating the connection between our invariants and other invariants described in the literature.

\section{The Isospin Formulation of the Problem} 
\label{Section: The Isospin Formulation of the Problem} 

\subsection{Mass and Flavor Isospin Operators} 
\label{Subsection: Mass and Flavor Isospin Operators} 
 
In this paper, we consider the mixing between two neutrino flavors. Without loss of generality, we can take one of these flavors to be $\nu_{e}$ and the other to be an orthogonal flavor state that we denote by $\nu_{x}$. In other words, $\nu_{x}$ can be either $\nu_{\mu}$ or $\nu_{\tau}$ or a normalized combination of them. We denote the fermion operator for an $\alpha$ flavor neutrino with momentum $\mathbf{p}$ by $a_{\alpha}\left(\mathbf{p}\right)$ where $\alpha=e,x$\footnote{In general, additional quantum numbers are needed in order to distinguish neutrinos with the same momentum. But to keep our formulas simple we do not explicitly include additional quantum numbers in our notation. Instead, one can view $\mathbf{p}$ as a multiple index like $(\mathbf{p},s_1,s_2,\dots)$.}.
The global rotation  
\begin{eqnarray} 
\label{Flavor to Mass Transformation} 
a_{e}(\mathbf{p}) & = &  \cos\theta\: a_{1}(\mathbf{p})+\sin\theta\: a_{2}(\mathbf{p})\\ 
a_{x}(\mathbf{p}) & = & -\sin\theta\: a_{1}(\mathbf{p})+\cos\theta\: a_{2}(\mathbf{p})\nonumber 
\end{eqnarray} 
relates them to the fermion operators $a_i\left(\mathbf{p}\right)$ for the corresponding mass eigenstates $\nu_{i}$ for $i=1,2$.  
 
Following the earlier literature, we introduce the \emph{flavor isospin operators}    
\begin{equation} 
{J}_{\mathbf{p}}^{+}= a_{e}^{\dagger}(\mathbf{p})a_{x}(\mathbf{p})~,\qquad 
{J}_{\mathbf{p}}^{-}= a_{x}^{\dagger}(\mathbf{p})a_{e}(\mathbf{p})~,\qquad 
{J}_{\mathbf{p}}^z=\frac{1}{2}\left(a_{e}^{\dagger}(\mathbf{p})a_{e}(\mathbf{p})-a_{x}^{\dagger}(\mathbf{p})a_{x}(\mathbf{p})\right)~, 
\label{Flavor Isospin Operators} 
\end{equation} 
which obey the usual $SU(2)$ commutation relations    
\begin{equation} 
[{J}_{\mathbf{p}}^{+},{J}_{\mathbf{q}}^{-}]=2\delta_{\mathbf{p}\mathbf{q}}{J}_{\mathbf{p}}^z~,\qquad 
[{J}_{\mathbf{p}}^z,{J}_{\mathbf{q}}^{\pm}]=\pm\delta_{\mathbf{p}\mathbf{q}}{J}_{\mathbf{p}}^{\pm}~, 
\label{Flavor Isospin Algebra} 
\end{equation} 
such that we have as many orthogonal $SU(2)$ \emph{flavor isospin algebras} as the number of neutrinos. It follows from the definitions given in Eq. (\ref{Flavor Isospin Operators}) that each flavor isospin algebra is realized in the spin-$1/2$ representation and that the electron neutrino is taken to be isospin up. We note, however, that sometimes the opposite convention for the isospin doublet is used (see, for example, Ref. \cite{Balantekin:2006tg}). 
 
We introduce the following summation convention for all quantities labeled by neutrino momentum $\mathbf{p}$:  
\begin{equation} 
A_{\omega}\equiv\sum_{|\mathbf{p}|=p}A_{\mathbf{p}} \qquad \mbox{and} \qquad 
A\equiv\sum_{\omega} A_{\omega}~. 
\label{Totals} 
\end{equation}
Here $\omega$ is the vacuum oscillation frequency for a neutrino with energy $p$. It is given by 
\begin{equation}
\label{Define w}
\omega=\frac{\delta m^2}{2p}
\end{equation} 
with $\delta m^2=m_2^2-m_1^2$ where $m_i$ is the mass of the state $\nu_i$. In the case of flavor isospin operators, for example, $\vec{J}_\omega$ represents the total flavor isospin operator of all neutrinos with the same vacuum oscillation frequency $\omega$ whereas $\vec{J}$ represents the total flavor isospin operator of all neutrinos. Since the operators $\vec{J}_{\omega}$ and $\vec{J}$ are sums of individual $SU(2)$ operators, their components also obey the $SU(2)$ commutation relations. The corresponding total flavor isospin quantum numbers can take several values. Note that in Eq. (\ref{Totals}) and in all subsequent equations sums are to be taken over all occupied neutrino states. Also note that we use boldface letters to indicate vectors in momentum space (e.g. $\mathbf{p}$) and arrows to indicate vectors in flavor space (e.g. $\vec{J}$). Throughout this paper, we use both the Cartesian basis $(x,y,z)$ and the cylindrical basis $(+,-,z)$ for vectors in flavor space. These two bases are related by 
\begin{equation}
J_\pm=J_x \pm iJ_y~. 
\end{equation}  
 
One can similarly introduce the \emph{mass isospin operators} 
\begin{equation} 
{\mathcal{J}}_{\mathbf{p}}^{+}= a_{1}^{\dagger}(\mathbf{p})a_{2}(\mathbf{p})~,\qquad 
{\mathcal{J}}_{\mathbf{p}}^{-}= a_{2}^{\dagger}(\mathbf{p})a_{1}(\mathbf{p})~,\qquad 
{\mathcal{J}}_{\mathbf{p}}^z=\frac{1}{2}\left(a_{1}^{\dagger}(\mathbf{p})a_{1}(\mathbf{p})-a_{2}^{\dagger}(\mathbf{p})a_{2}(\mathbf{p})\right)~, 
\label{Mass Isospin Operators} 
\end{equation} 
which also obey the $SU\left(2\right)$ commutation relations given in Eq. (\ref{Flavor Isospin Algebra}).   
As can be seen from the definitions given in Eq. (\ref{Mass Isospin Operators}), we take the mass eigenstate $\nu_1$ to be isospin up. The total mass isospin operators are defined as in Eq. (\ref{Totals}). 
 
Eqs. (\ref{Flavor to Mass Transformation}) imply that the particle operators in the mass and flavor bases are related by a unitary transformation. The most general unitary transformation operator in flavor space can be written as
\begin{eqnarray} 
{U_{\mathbb{p}}}=e^{\sum_{\mathbf{p}} z_{\mathbf{p}}{\mathcal{J}}_{\mathbf{p}}^+}\;e^{\sum_{\mathbf{p}} \ln\left(1+\left|z_{\mathbf{p}}\right|^2\right){\mathcal{J}}_{\mathbf{p}}^z}\;e^{-\sum_{\mathbf{p}} z_{\mathbf{p}}{\mathcal{J}}_{\mathbf{p}}^-} 
\label{General Unitary Operator} 
\end{eqnarray} 
where $z_{\mathbf{p}}=e^{i\delta_{\mathbf{p}}}\tan\theta_{\mathbf{p}}$ is a complex parameter. The effect of such a general transformation on particle operators is
\begin{eqnarray} 
\label{General Unitary Transformation} 
{U_{\mathbb{p}}}^{\dagger} a_1(\mathbf{p}){U_{\mathbb{p}}}^{}  
&=& \cos{\theta_{\mathbf{p}}} \; a_1(\mathbf{p}) -e^{i\delta_{\mathbf{p}}}\sin{\theta_{\mathbf{p}}} \; a_2(\mathbf{p}) \\ 
{U_{\mathbb{p}}}^{\dagger} a_2(\mathbf{p}){U_{\mathbb{p}}}^{}  
&=& e^{-i\delta_{\mathbf{p}}}\sin{\theta_{\mathbf{p}}} \; a_1(\mathbf{p})+\cos{\theta_{\mathbf{p}}} \; a_2(\mathbf{p})\nonumber 
\end{eqnarray} 
i.e., each momentum mode is rotated by a different angle and acquires a different (relative) phase. The transformation from mass to flavor basis given in Eqs. (\ref{Flavor to Mass Transformation}) is a special case of Eqs. (\ref{General Unitary Transformation}) for which the rotation angle is equal to the vacuum mixing angle and the phase is zero for all momentum modes. In other words, we can write   
\begin{equation} 
a_e({\mathbf{p}})={U}^{\dagger}a_1({\mathbf{p}}){U}   
\qquad\mbox{and}\qquad 
a_x({\mathbf{p}})={U}^{\dagger}a_2({\mathbf{p}}){U}~,
\label{Unitary Transformation of Particle Operators} 
\end{equation} 
where 
\begin{equation}
\label{Unitary Operator} 
{U}= 
e^{z{\mathcal{J}}^+}\;e^{\ln\left(1+\left|z\right|^2\right){\mathcal{J}}^z}\;e^{-z{\mathcal{J}}^-}  
\end{equation} 
with $z_\mathbf{p}=z=\tan\theta$ for all $\mathbf{p}$. Note that we use the subscript $\mathbb{p}$ in denoting the operator $U_\mathbb{p}$ defined in Eq. (\ref{General Unitary Operator}) in order to emphasize that it imposes a different rotation on each momentum mode. Although the operator $U$ is only a special case of $U_{\mathbb{p}}$, it carries no such index because it induces the same transformation on all momentum modes. Also note that the rotation parameter $z_\mathbf{p}$ is not subject the summation rule introduced in Eq. (\ref{Totals}). Instead, its index simply indicates its dependence on momentum.  

The operator $U_\mathbb{p}$ given in Eq. (\ref{General Unitary Operator}) also induces a transformation on the isospin operators as follows:  
\begin{eqnarray} 
U^\dagger_\mathbb{p}\mathcal{J}_{\mathbf{p}}^z U_\mathbb{p} &=& 
\cos2\theta_\mathbf{p}\;\mathcal{J}_{\mathbf{p}}^z+\frac{1}{2} e^{i\delta_\mathbf{p}}\sin2\theta_\mathbf{p} \; \mathcal{J}_{\mathbf{p}}^+ + \frac{1}{2} e^{-i\delta_\mathbf{p}}\sin2\theta_\mathbf{p} \; \mathcal{J}_{\mathbf{p}}^-~,
\nonumber\\ 
U^\dagger_\mathbb{p}\mathcal{J}_{\mathbf{p}}^+ U_\mathbb{p} &=& 
\cos^2 \theta_\mathbf{p} \mathcal{J}_{\mathbf{p}}^+ -e^{-i\delta_\mathbf{p}}\sin2\theta_\mathbf{p} \mathcal{J}_{\mathbf{p}}^z -e^{-2i\delta_\mathbf{p}}\sin^2 \theta_\mathbf{p} \mathcal{J}_{\mathbf{p}}^-~, 
\label{General Transformation}\\ 
U^\dagger_\mathbb{p}\mathcal{J}_{\mathbf{p}}^- U_\mathbb{p} &=& 
\cos^2 \theta_\mathbf{p} \mathcal{J}_{\mathbf{p}}^- -e^{i\delta_\mathbf{p}}\sin2\theta_\mathbf{p} \mathcal{J}_{\mathbf{p}}^z -e^{2i\delta_\mathbf{p}}\sin^2 \theta_\mathbf{p} \mathcal{J}_{\mathbf{p}}^+~.
\nonumber
\end{eqnarray}  
In particular, mass and flavor isospin operators are related by
\begin{equation} 
\vec{J}_{\mathbf{p}}={U}^{\dagger}\vec{\mathcal{J}}_{\mathbf{p}}{U}~ 
\label{Unitary Transformation} 
\end{equation} 
which implies
\begin{eqnarray} 
{J}_{\mathbf{p}}^z &=& 
\cos2\theta\;\mathcal{J}_{\mathbf{p}}^z+\sin2\theta\;\frac{\mathcal{J}_{\mathbf{p}}^+ +\mathcal{J}_{\mathbf{p}}^-}{2}~, 
\nonumber\\ 
{J}_{\mathbf{p}}^+ &=& 
\cos^2 \theta \mathcal{J}_{\mathbf{p}}^+ -\sin2\theta \mathcal{J}_{\mathbf{p}}^z -\sin^2 \theta \mathcal{J}_{\mathbf{p}}^-~, 
\label{Transformation}\\ 
{J}_{\mathbf{p}}^- &=& 
\cos^2 \theta \mathcal{J}_{\mathbf{p}}^- -\sin2\theta \mathcal{J}_{\mathbf{p}}^z -\sin^2 \theta \mathcal{J}_{\mathbf{p}}^+~. 
\nonumber
\end{eqnarray} 
The inverse transformation from flavor to mass basis can be found by substituting $-\theta$ in place of $\theta$ in Eq. (\ref{Transformation}). Note that Eqs. (\ref{Unitary Operator}) and (\ref{Unitary Transformation}) follow from the fact that the total isospin operator $\vec{J}$ is the generator of global rotations in flavor space.

\subsection{The Hamiltonian and the Quantum Invariants} 
\label{Subsection: The Hamiltonian and the Quantum Invariants} 
 
The many-body Hamiltonian describing the vacuum oscillations of a group of neutrinos is given by 
\begin{equation} 
{H}_{\nu}=\sum_{\mathbf{p}}\left(\frac{m_{1}^{2}}{2p}a_{1}^{\dagger}(\mathbf{p})a_{1}(\mathbf{p})+\frac{m_{2}^{2}}{2p}a_{2}^{\dagger}(\mathbf{p})a_{2}(\mathbf{p})\right)~. 
\label{Vacuum Oscillation Hamiltonian} 
\end{equation} 
One can write this Hamiltonian in terms of the neutrino isospin operators defined in Section \ref{Subsection: Mass and Flavor Isospin Operators}. First note that since the neutrinos only exchange their momenta with the forward scattering, the total number of neutrinos in each momentum mode is constant. As a result, the term  
\begin{equation} 
\sum_{\mathbf{p}}\frac{m_{1}^{2}+m_{2}^{2}}{4p}(a_{1}^{\dagger}(\mathbf{p})a_{1}(\mathbf{p})+a_{2}^{\dagger}(\mathbf{p})a_{2}(\mathbf{p})) 
\end{equation} 
is proportional to identity for a given number of particles and can be subtracted from the Hamiltonian without any consequences. Subtracting this term from Eq. (\ref{Vacuum Oscillation Hamiltonian}) and using the definitions given in Eqs. (\ref{Define w}) and (\ref{Mass Isospin Operators}) together with the transformation given in Eqs. (\ref{Transformation}) one finds 
\begin{equation} 
\label{Vacuum Oscillations} 
{H}_\nu=\sum_{\omega}\frac{\delta m^2}{2p} \vec{B}\cdot\vec{J}_{\omega}~. 
\end{equation} 
Here $\vec{B}$ is the unit vector which points in the \emph{mass direction}. Its components are given by 
\begin{equation} 
\vec{B}=(0,0,-1)_{\mbox{\tiny mass}}=(\sin2\theta,0,-\cos2\theta)_{\mbox{\tiny flavor}}~ 
\end{equation} 
in mass and flavor bases, respectively.
 
In writing the Hamiltonian which describes the self refraction of neutrinos, one should take into account the fact that there are two kinds of neutrino-neutrino scattering diagrams which add up coherently during the neutrino propagation. One of them is the forward scattering diagram in which there is no momentum transfer and both neutrinos remain in their original states after the scattering. These diagrams give rise to a diagonal refraction potential in the flavor basis, similar to the MSW potential. The other one is the exchange diagram in which the neutrinos exchange their states after the scattering. These diagrams give rise to a non-diagonal refraction potential in the flavor basis which is the source of the nonlinearity of the neutrino self refraction problem \cite{Pantaleone:1992eq,Pantaleone:1992xh,Sigl:1992fn}. Taking into account the contribution of both kinds of diagrams mentioned above, the effective many-body Hamiltonian describing the self refraction of a dense neutrino gas can be written as \cite{Sawyer:2005jk}  
\begin{eqnarray} 
{H}_{\nu\nu}=\frac{G_{F}}{\sqrt{2}V}\sum_{\mathbf{p}}\sum_{\mathbf{q}} 
(1-\cos\vartheta_{\mathbf{p}\mathbf{q}})  \!\!\!\!\!\!\!\!\!\!& \left[\right.a_{e}^{\dagger}(\mathbf{p})a_{e}(\mathbf{p})a_{e}^{\dagger}(\mathbf{q})a_{e}(\mathbf{q})+a_{x}^{\dagger}(\mathbf{p})a_{x}(\mathbf{p})a_{x}^{\dagger}(\mathbf{q})a_{x}(\mathbf{q})\nonumber \\ 
 & +a_{x}^{\dagger}(\mathbf{p})a_{e}(\mathbf{p})a_{e}^{\dagger}(\mathbf{q})a_{x}(\mathbf{q})+a_{e}^{\dagger}(\mathbf{p})a_{x}(\mathbf{p})a_{x}^{\dagger}(\mathbf{q})a_{e}(\mathbf{q}))\left.\right]~. 
\label{Self Interaction Hamiltonian} 
\end{eqnarray} 
Here $V$ is the quantization volume and $\vartheta_{\mathbf{p}\mathbf{q}}$ is the angle between the momentum modes $\mathbf{p}$ and $\mathbf{q}$. The factor $1-\cos\vartheta_{\mathbf{p}\mathbf{q}}$ guarantees that those neutrinos traveling in the same direction do not undergo scattering.  
 
One can write the neutrino self refraction Hamiltonian in terms of the neutrino isospin operators defined in Section \ref{Subsection: Mass and Flavor Isospin Operators} as follows: 
\begin{equation} 
\label{Self Interactions} 
{H}_{\nu\nu} 
=\frac{\sqrt{2}G_{F}}{V}\sum_{\mathbf{p},\mathbf{q}}\left(1-\cos\vartheta_{\mathbf{p}\mathbf{q}}\right)\vec{J}_{\mathbf{ p}}\cdot\vec{J}_{\mathbf{q}}~. 
\end{equation} 
Here, Eq. (\ref{Self Interactions}) can be directly obtained from Eq. (\ref{Self Interaction Hamiltonian}) by using the definitions given in Eq. (\ref{Flavor Isospin Operators}) and discarding those terms proportional to identity.  Eq. (\ref{Self Interactions}) tells us that in the language of flavor isospins the neutrino-neutrino interaction takes the form of a spin exchange interaction. This is an expected result because the effective Hamiltonian given in Eq. (\ref{Self Interaction Hamiltonian}) consists only of those neutrino-neutrino interactions in which neutrinos either keep or exchange their momenta.  
 
The flavor evolution of a dense neutrino gas (in the absence of any other background) is described by the sum of the vacuum oscillation term and self interaction term  
\begin{equation} 
{H}= \sum_{\omega}\omega\vec{B}\cdot\vec{J}_{\omega}+\mu\sum_{\mathbf{p},\mathbf{q}}\left(1-\cos\vartheta_{\mathbf{p}\mathbf{q}}\right)\vec{J}_{\mathbf{p}}\cdot\vec{J}_{\mathbf{q}}~. 
\end{equation} 
Here we defined
\begin{equation} 
\mu=\frac{\sqrt{2}G_{F}}{V}~. 
\label{Define mu} 
\end{equation}
In this study, we adopt what is commonly referred to as the ``single angle approximation,'' i.e., we will assume that the term involving $\cos\vartheta_{\mathbf{p}\mathbf{q}}$ in the Hamiltonian averages to zero so that the neutrinos traveling in different directions which are otherwise identical undergo the same flavor evolution. With this simplification, the total Hamiltonian becomes  
\begin{subequations}
\label{Hamiltonian} 
\begin{eqnarray} 
{H}&=& -\sum_{\omega}\omega\mathcal{J}^z_{\omega} +\mu\vec{\mathcal{J}}\cdot\vec{\mathcal{J}}  
\label{Hamiltonian in Mass Basis}\\ 
&=& \sum_{\omega}\omega\vec{B}\cdot\vec{J}_{\omega} +\mu\vec{J}\cdot\vec{J}
\label{Hamiltonian in Flavor Basis}~. 
\end{eqnarray}
\end{subequations}
Note that the self interaction term has the same form in both mass and flavor bases because these two bases are related by a \emph{global} rotation as described in Eq. (\ref{Unitary Transformation}) which leaves \emph{all} scaler products invariant.
 
It has been pointed out by many authors that the Hamiltonian given in Eq. (\ref{Hamiltonian}) is analogous to the Hamiltonian of an \emph{interacting spin system}, i.e., a group of spins interacting with a position dependent \emph{external magnetic field} and with each other via spin exchange interaction. In the mass basis the external magnetic field points in the $-z$ direction whereas in flavor basis it points in the direction of the unit vector $\vec{B}=(\sin2\theta,0,-\cos2\theta)$. This analogy makes it easier to see that the length of each isospin is conserved:  
\begin{equation} 
\label{Conservation of Spin Length} 
{L}_{\omega}=\vec{J}_{\omega}\cdot\vec{J}_{\omega} 
\qquad\qquad\left[ {H}, {L}_{\omega}\right]=0. 
\end{equation} 
Similarly, the total isospin component in the direction of the external magnetic field is also conserved:  
\begin{equation} 
\label{Conservation of Q_3} 
{C}_0=\vec{B}\cdot\vec{J} 
\qquad\qquad\left[{H},{C}_0\right]=0~. 
\end{equation} 
 
It is worth noting that the Hamiltonian in Eq. (\ref{Hamiltonian}) also appears in connection with the pairing problem. For example, in the BCS theory of superconductivity \cite{Bardeen:1957mv} the (reduced) pairing interaction between the valance electrons is described by the Hamiltonian\footnote{The most general (i.e., non reduced) pairing Hamiltonian has the form  
\begin{equation*} 
{H}_{\mbox{\tiny BCS}}=\sum_{k}2\epsilon_{k}{t}_{k}^z-G\sum_{k,k^\prime} c_{kk^\prime} t^+_k t^-_{k^\prime}~, 
\end{equation*} 
where the dimensionless coefficients $c_{kk^\prime}$ lead to a state dependent pairing strength similar to the coefficient $1-\cos\vartheta_{\mathbf{p}\mathbf{q}}$ in the neutrino Hamiltonian. 
} 
\begin{equation} 
{H}_{\mbox{\tiny BCS}}=\sum_{k}2\epsilon_{k}{t}_{k}^z-G{T}^+{T}^-~. 
\label{BCS Hamiltonian} 
\end{equation} 
Here $G$ is a constant which represents the strength of the pairing interaction. Within the context of the BCS model, it is assumed that $G>0$ leading to an attractive pairing and making the formation of Cooper pairs energetically favorable. $\epsilon_k$ are the degenerate single particle energy levels which can be occupied by pairs of spin-up and spin-down electrons (i.e., the Cooper pairs). The operators 
\begin{equation} 
{t}_k^+=c_{k\uparrow}^\dagger c_{k\downarrow}^\dagger~, \qquad  
{t}_k^-=c_{k\downarrow} c_{k\uparrow}~ \qquad\mbox{and}\qquad 
{t}_k^z=\frac{1}{2}\left(c_{k\uparrow}^\dagger c_{k\uparrow}+c_{k\downarrow}^\dagger c_{k\downarrow}-1\right) 
\label{Quasi-spin Operators} 
\end{equation} 
are called \emph{quasi-spin operators} and they also obey the same $SU(2)$ commutation relations as given in Eq. (\ref{Flavor Isospin Algebra}). In the quasi-spin scheme, a single particle state $k$ has quasi-spin up if it is occupied by a pair and quasi-spin down if it is not. The operator  
\begin{equation} 
\vec{T}=\sum_k \vec{t}_k 
\end{equation} 
represents the total quasi-spin. As can be easily verified, the third component of the total quasi-spin is a constant of motion, i.e., 
\begin{equation} 
[{H}_{\mbox{\tiny BCS}},{T}^z]=0~, 
\label{Conservation of T3} 
\end{equation} 
which is analogous to the conservation of $C_0$ mentioned in Eq. (\ref{Conservation of Q_3}). From Eq. (\ref{Quasi-spin Operators}) we see that $2{T}^z$ is equal to the total number of electron pairs in the system. 
 
The Hamiltonians given in Eqs. (\ref{Hamiltonian in Mass Basis}) and (\ref{BCS Hamiltonian}) are very similar. In fact, if we make the substitutions $\omega\to 2\epsilon_k$ and $\vec{\mathcal{J}}_{\mathbf{p}}\to\vec{t}_k$ and keep in mind that the `missing' term $G{T}^z({T}^z-1)$ in Eq. (\ref{BCS Hamiltonian}) is proportional to identity for a given number of pairs (i.e., has no effect on the evolution), then we see that the two models are mathematically identical up to an overall minus sign (remember that $\mu$ is by definition a positive quantity). The correspondence between these two models can be summarized as follows: Single particle states with energy $\epsilon_k$ in the BCS model correspond to the neutrino states with oscillation frequency $\omega$. Both sets of states are multiply degenerate. A state $\epsilon_k$ which is occupied (respectively, unoccupied) by a pair in the BCS model corresponds to a neutrino state with oscillation frequency $\omega$ occupied by a neutrino in $\nu_1$ (respectively, $\nu_2$) mass eigenstate in the neutrino model.  
 
It was first shown by Richardson in 1963 that the BCS Hamiltonian given in Eq. (\ref{BCS Hamiltonian}) can be diagonalized analytically with the method of algebraic Bethe ansatz \cite{Bethe,Richardson1}. Later work by Gaudin \cite{Gaudin1, Gaudin2} and others \cite{Cambiaggio,Dukelsky:2004re,Sierra:2001cx} revealed that the integrability of the Hamiltonian in Eq. (\ref{BCS Hamiltonian}) derives from the existence of a set of \emph{constants of motion} or \emph{quantum invariants}. These results can be easily carried over from BCS model to the neutrino model. For example, it can be shown that the neutrino Hamiltonian given in Eq. (\ref{Hamiltonian}) has the following constants of motion:  
\begin{equation} 
\label{Invariants} 
{h}_{\omega}=\vec{B}\cdot\vec{J}_{\omega}+2\mu\sum_{\omega^\prime\left(\neq\omega\right)}\frac{\vec{J}_{\omega}\cdot\vec{J}_{\omega^\prime}}{\omega-{\omega^\prime}}~. 
\end{equation} 
The operators given in Eq. (\ref{Invariants}) are known as the \emph{Gaudin magnet Hamiltonians}. It is straightforward to show that these operators commute with one another and with the Hamiltonian given in Eq. (\ref{Hamiltonian}), i.e.,  
\begin{equation} 
\left[{h}_{\omega},{h}_{\omega^\prime}\right]=0\qquad\mbox{and}\qquad\left[{H},{h}_{\omega}\right]=0 
\label{Conservation of GMH} 
\end{equation} 
is satisfied for every $\omega$ and $\omega^\prime$. Note that the invariants given in Eq. (\ref{Invariants}) are independent from one another and from the invariants mentioned in Eq. (\ref{Conservation of Spin Length}). However, the invariant ${C}_0$ mentioned in Eq. (\ref{Conservation of Q_3}) is the sum of the Gaudin magnet Hamiltonians 
\begin{equation} 
\label{Q^z Combination} 
{C}_0=\sum_{\omega}{h}_{\omega}  
\end{equation} 
and thus is not an independent invariant. The Hamiltonian itself is also a linear combination of the invariants given in Eqs. (\ref{Conservation of Spin Length}) and (\ref{Invariants}):   
\begin{equation}
\label{C1} 
{H}=\sum_{\omega}\omega{h}_{\omega}+\sum_{\omega} {L}_{\omega}~. 
\end{equation} 
This tells us that if the neutrinos occupy $\Omega$ different energy modes, then the system has $2\Omega$ independent constants of motion given in Eqs. (\ref{Conservation of Spin Length}) and (\ref{Invariants}). These invariants can be expressed in several alternative forms which may be useful in different applications. In Appendix \ref{Appendix Invariants}, we present two alternative ways of writing down the invariants. 
 
From the definitions given in Eq. (\ref{Mass Isospin Operators}), one can see that the constant $C_0$ can be written as 
\begin{equation} 
C_0=-{\mathcal{J}}^z=\frac{\hat{N}_{2}-\hat{N}_{1}}{2}~ 
\label{Q0} 
\end{equation} 
where 
\begin{equation} 
\hat{N}_1=\sum_{\mathbf{p}} a_{1}^{\dagger}(\mathbf{p})a_{1}(\mathbf{p}) 
\qquad \mbox{and} \qquad 
\hat{N}_2=\sum_{\mathbf{p}} a_{2}^{\dagger}(\mathbf{p})a_{2}(\mathbf{p}) 
\end{equation} 
denote the total particle number operators for the mass eigenstates $\nu_1$ and $\nu_2$, respectively. Note that in this paper we denote the particle \emph{number operators} with hats (e.g., $\hat{N}_i$) and the corresponding particle \emph{numbers} without hats (e.g., ${N}_i$). Together with the fact that the total number of neutrinos $N=N_1+N_2$ is constant, the conservation of $C_0$ mentioned in Eq. (\ref{Conservation of Q_3}) ensures that the Hamiltonian preserves the number of neutrinos in each mass eigenstate, i.e., we have 
\begin{equation} 
\left[{H}, \hat{N}_i \right]=0~, 
\label{Particle Number Conservation} 
\end{equation}   
for $i=1,2$.  
 
Although $C_0$, which is a combination of the invariants ${h}_{\omega}$, can be simply expressed in terms of particle number operators as in Eq. (\ref{Q0}), this is an exceptional situation. Apart from this particular case, the invariants ${h}_{\omega}$ or their combinations cannot be written in terms of the particle number operators. This is evident from the fact that the terms $\vec{J}_{\omega}\cdot\vec{J}_{\omega^\prime}$ in Eq. (\ref{Invariants}) are diagonal in neither the mass basis nor the flavor basis and that these terms disappear only in the particular combination given in Eq. (\ref{Q^z Combination}).

\section{Diagonalization of the Hamiltonian} 
\label{Section: Diagonalization of the Hamiltonian} 

The existence of constants of motion is a manifestation of exact solvability of the pairing Hamiltonian given in Eq. (\ref{Hamiltonian}). As mentioned earlier in previous sections, the pioneering work in this direction was that of Richardson who showed that the application of the Bethe ansatz method to the pairing Hamiltonian yields its exact eigenstates and eigenvalues in an analytical way. This exact solvability of the pairing Hamiltonian has been studied and exploited extensively thereafter both in the context of the nuclear shell model and in the context of BCS theory. In this section, we will review this procedure for a self interacting neutrino gas.  

The Bethe ansatz method, if applicable to a problem, usually reveals a general and simple \emph{functional form} for the many-body eigenstates. In the case of the self interacting neutrino gas, the many-body eigenstates turn out to resemble those of an harmonic oscillator in that they can be obtained by repeated application of a step operator on a lowest weight state. The crucial difference is that in the neutrino case the step operator is parametrized by a complex number. This parameter takes on different values at each step and should be determined by solving the equations of Bethe ansatz as described in more detail below. The number of equations that one is required to solve in order to find the \emph{full} spectrum  \emph{exactly} is of the order of the number of particles. For a dense neutrino gas, this is clearly unfeasible. However, if an approximate method of determining the Bethe ansatz variables can be found, then the simple step operator form of the eigenstates allows one to write down an (approximate) evolution operator for the system. Alternatively, one can consider a relatively small number of neutrinos occupying a very small volume so as to yield a large density. Mimicking the conditions in a dense environment in this way allows one to explore many-body physics with the exact eigenstates. One can then increase the number of neutrinos and the volume that they occupy keeping the density constant. The resulting limit of the Bethe ansatz equations have been studied extensively in the context of the BCS model (see, for example, Refs. \cite{Dukelsky:2004re,Richardson:1966zza,Richardson Large N,Gaudin Electrostatic Analogy,Roman:2002dh,Amico:2002hq} and the references therein) and in connection with matrix models that appear in 2D gravity \cite{Jurco:2003pv}. This limit will be reviewed in Section \ref{Subsection: Electrostatic Analogy}.    

We would like to note that a study of the general functional form of the eigenstates as revealed by the Bethe ansatz method can itself give insight into the collective behavior of the system. For example, it was shown in Ref. \cite{yuzb} that the expectation value of the step operator in the effective one-particle approximation to the pairing problem is the generating function for the canonical variables of the system and that the exact analytical solutions of the corresponding RPA equations of motion can be obtained in terms of these variables.  
 
\subsection{The Bethe Ansatz Method} 
\label{Subsection: The Bethe Ansatz Method} 
 
The eigenstates of the neutrino Hamiltonian given in Eq. (\ref{Hamiltonian}) can be easily found in the two opposite limits of the parameter $\mu$ defined in Eq. (\ref{Define mu}).  
 
\begin{itemize} 
\item 
{\bf As $\mu$ approaches to zero}, neutrinos occupy a larger and larger volume and neutrino-neutrino scattering becomes negligible. In this limit, the Hamiltonian consists only of vacuum oscillations given in Eq. (\ref{Vacuum Oscillations}). The eigenstates in this limit are simply the tensor products of mass eigenstates for individual neutrinos. Speaking in terms of the interacting spin system analogy, in this limit the spins interact only with the external magnetic field which is in the $-z$ direction in mass basis. The eigenstates are those in which each spin is either aligned or anti-aligned with the external magnetic field, i.e.,  
\begin{equation} 
|\nu_1 \; \nu_1 \; \nu_1 \; \dots \rangle, \quad |\nu_2 \; \nu_1 \; \nu_1 \; \dots \rangle, \quad |\nu_1 \; \nu_2 \; \nu_1 \; \dots \rangle, \quad |\nu_1 \; \nu_1 \; \nu_2 \; \dots \rangle, \quad \dots 
\end{equation} 
 
\item 
{\bf As $\mu$ approaches to infinity}, neutrinos are crowded into a smaller and smaller volume and eventually the neutrino-neutrino scattering term becomes dominant. In this limit, one can ignore the vacuum oscillations and write the Hamiltonian as  
\begin{equation} 
\label{Large mu Limit of the Hamiltonian} 
{H}_\infty\equiv\lim_{\mu\to\infty}{H}=\mu\vec{J}\cdot\vec{J}~. 
\end{equation} 
The \emph{total isospin quantum number} $j$ is a scalar quantity so that we can use it without any references to mass or flavor bases. Note that $j$ can take several values starting from $0$ or $1/2$ (depending on whether we have an even or odd number of neutrinos) up to $j_{\mbox{\tiny max}}=N/2$ where $N$ is the total number of neutrinos. In the limit where $\mu\to\infty$, both the total mass isospin states $ |j , m\rangle_m$ and the total flavor isospin states $ |j , m\rangle_f$ are eigenstates of the Hamiltonian with the same energy, i.e.,  
\begin{equation} 
{H}_\infty |j , m\rangle_m =\mu j(j+1) |j , m\rangle_m 
\qquad \mbox{and} \qquad 
{H}_\infty |j , m\rangle_f =\mu j(j+1) |j , m\rangle_f~. 
\label{Large mu Limit of the Eigenstates} 
\end{equation}  
It is clear from Eq. (\ref{Unitary Transformation}) that these two sets of states are related by 
\begin{equation} 
|j , m\rangle_f={U}^\dagger|j , m\rangle_m~. 
\label{State Transformation} 
\end{equation} 
Since the operator ${U}$ involves only the total isospin operators (see Eq. (\ref{Unitary Operator})) it cannot change the representation in which a state lives, i.e., it preserves the value of $j$ in accordance with the above arguments.  
 
To illustrate the use of Eq. (\ref{State Transformation}), let us consider a state in which all neutrinos are $\nu_e$. Since this corresponds to having all flavor isospins up, this state is the highest weight state of the total flavor isospin algebra, i.e.,  
\begin{equation}  
|\nu_e \; \nu_e \; \nu_e \; \dots \rangle = |j_{\mbox{\tiny max}} , j_{\mbox{\tiny max}}\rangle_f~. 
\end{equation} 
This state can be converted to mass basis using Eqs. (\ref{Unitary Operator}) and (\ref{State Transformation}). The result is 
\begin{eqnarray} 
|j_{\mbox{\tiny max}} , j_{\mbox{\tiny max}}\rangle_f &=&{U}^\dagger|j_{\mbox{\tiny max}} , j_{\mbox{\tiny max}}\rangle_m\\ 
&=&\sum_{m=-j_{\mbox{\tiny max}}}^{j_{\mbox{\tiny max}}} (\cos\theta)^{j_{\mbox{\tiny max}}-m} (\sin\theta)^{j_{\mbox{\tiny max}}+m} \sqrt{\frac{(2j_{\mbox{\tiny max}})!}{(j_{\mbox{\tiny max}}-m)! (j_{\mbox{\tiny max}}+m)!}}\; \nonumber 
|j_{\mbox{\tiny max}} , m\rangle_m~. 
\end{eqnarray} 
We see that the state $|j_{\mbox{\tiny max}} , j_{\mbox{\tiny max}}\rangle_f$ is a linear combination of the states $|j_{\mbox{\tiny max}} , m\rangle_m$, which all live in the $j=j_{\mbox{\tiny max}}$ representation and have the same energy in the limit where $\mu\to\infty$.  
\end{itemize} 
 
We next present the eigenstates and eigenvalues of the collective neutrino oscillation Hamiltonian given in Eq. (\ref{Hamiltonian}) away from those two limits, i.e., for $0<\mu<\infty$. We  begin by noting that Eq. (\ref{Particle Number Conservation}) implies that all eigenstates of the collective oscillation Hamiltonian are also eigenstates of the number operators $\hat{N}_1$ and $\hat{N}_2$, i.e., each many-body eigenstate has a definite number of neutrinos in $\nu_1$ and $\nu_2$ states.  
 
In general, the total isospin states $|j , m\rangle_m$ and $|j , m\rangle_f$ mentioned in Eq. (\ref{Large mu Limit of the Eigenstates}) are no longer eigenstates of the Hamiltonian for a finite value of $\mu$ but the highest and lowest weight states $|j , \pm j \rangle_m$ of the total mass isospin continue to be eigenstates. Let us consider, for example, the states in which all neutrinos occupy the same mass eigenstate, i.e., they are all $\nu_1$ or all $\nu_2$. Speaking in the language of isospin, this is equivalent to having all mass isospins up or all mass isospins down, respectively. These situations respectively correspond to the total isospin states $|j_{\mbox{\tiny max}},j_{\mbox{\tiny max}}\rangle_m$ and $|j_{\mbox{\tiny max}},-j_{\mbox{\tiny max}}\rangle_m$: 
\begin{equation} 
\label{Define Phi j_max} 
|j_{\mbox{\tiny max}},j_{\mbox{\tiny max}}\rangle_m= \prod_{\mathbf{p}}a_{1}^{\dagger}(\mathbf{p})\;|0\rangle 
\qquad \mbox{and} \qquad 
|j_{\mbox{\tiny max}},-j_{\mbox{\tiny max}}\rangle_m= \prod_{\mathbf{p}}a_{2}^{\dagger}(\mathbf{p})\;|0\rangle~. 
\end{equation} 
One can easily show that these states are eigenstates of the Hamiltonian given in Eq. (\ref{Hamiltonian}) with the respective energies  
\begin{equation} 
E_{(+j_{\mbox{\tiny max}})}=-\sum_{\omega}\frac{\omega N_{\omega}}{2}+\mu j_{\mbox{\tiny max}}\left(j_{\mbox{\tiny max}}+1\right) 
\qquad\mbox{and}\qquad  
E_{(-j_{\mbox{\tiny max}})}=\sum_{\omega}\frac{\omega N_{\omega}}{2}+\mu j_{\mbox{\tiny max}}\left(j_{\mbox{\tiny max}}+1\right)~. 
\label{Zero Energies j_max} 
\end{equation} 
Here we denote the total number of neutrinos in an individual energy mode by $N_{\omega}$. The other highest and lowest weight states $|j ,\pm j \rangle_m$ of the total mass isospin are also eigenstates of the Hamiltonian. To see this, let us first consider the total isospin quantum number $j_{\omega}$ of an energy mode which can take several values from $0$ or $1/2$ (depending whether we have an even or odd number of neutrinos in this mode) up to $N_{\omega}/2$, possibly with many multiplicities. For each energy mode, we can write 
\begin{equation} 
\label{States of Energy Modes} 
{\mathcal{J}}^z_{\omega} |j_{\omega} , \pm j_{\omega}\rangle_m=\pm j_{\omega} |j_{\omega} , \pm j_{\omega}\rangle_m~. 
\end{equation} 
Therefore the states 
\begin{eqnarray} 
|j , j \rangle_m &\equiv& |j_{\omega_1}, j_{\omega_1}\rangle_m \otimes |j_{\omega_2}, j_{\omega_2}\rangle_m \otimes \dots \otimes |j_{\omega_\Omega}, j_{\omega_\Omega}\rangle_m 
\label{Define Phi j}\\ 
|j , -j \rangle_m &\equiv& |j_{\omega_1}, -j_{\omega_1}\rangle_m \otimes |j_{\omega_2},-j_{\omega_2}\rangle_m \otimes \dots \otimes |j_{\omega_\Omega}, -j_{\omega_\Omega}\rangle_m\nonumber 
\end{eqnarray} 
are simultaneous eigenstates of all ${\mathcal{J}}^z_{\omega}$ (see Eq. (\ref{States of Energy Modes})) as well as the quadratic operator $\vec{\mathcal{J}}\cdot\vec{\mathcal{J}}$ (see Eq. (\ref{Large mu Limit of the Eigenstates})). As a result, the states given in Eqs. (\ref{Define Phi j}) are eigenstates of the Hamiltonian (\ref{Hamiltonian}) for all values of $\mu$ with the respective energies 
\begin{equation} 
E_{(+j)}=-\sum_{\omega}\omega j_{\omega}+\mu j\left(j+1\right) 
\qquad\mbox{and}\qquad  
E_{(-j)}=\sum_{\omega}\omega j_{\omega}+\mu j\left(j+1\right)~. 
\label{Zero Energies j} 
\end{equation} 
In Eqs. (\ref{Define Phi j}), we used $\Omega$ to denote the number of different energy modes. In this equation the total angular momentum $j=j_{p_1}+j_{p_2}+\dots+j_{p_\Omega}$ may come with many multiplicities because in general more than one combination of $j_{p_1}, j_{p_2},\dots, j_{p_\Omega}$ may correspond to the same sum $j$. Such multiplicities are inherent in the addition of angular momenta and one usually introduces additional quantum numbers to distinguish the resulting degenerate total angular momentum states. However, in this paper we avoid introducing such quantum numbers to keep the formulas readable. Since the energies given in Eq. (\ref{Zero Energies j}) depend only on $j_{\omega}$ and $j$, this omission does not lead to confusion in what follows. Finally note that Eqs. (\ref{Define Phi j_max}) and (\ref{Zero Energies j_max}) correspond the \emph{unique} special case of Eqs. (\ref{Define Phi j}) and (\ref{Zero Energies j}) in which we have $j_{\omega}=N_{\omega}/2$ for all energy modes and thus $j=j_{\mbox{\tiny max}}$.  
 
The eigenstates of the Hamiltonian other than the highest and lowest weight states $|j , \pm j \rangle_m$ can be found with the Bethe ansatz technique. This method is based on writing down a trial state depending on a set of unknown parameters which are called \emph{Bethe ansatz variables}. The values of these parameters are determined subject to the requirement that the state be an eigenstate of the Hamiltonian. For the particular problem at hand, the trial state is formed with the help of the \emph{Gaudin operators}  
\begin{equation} 
\label{Gaudin Operators} 
\mathcal{Q}^\pm(\xi) = \sum_{\omega} \frac{{\mathcal{J}}^\pm_{\omega}}{\omega-\xi}~. 
\end{equation} 
Here $\xi$ is a complex parameter which will later play the role of a Bethe ansatz variable. In order to demonstrate the method of Bethe ansatz, let us consider the state $|j , -j \rangle_m$ defined in Eq. (\ref{Define Phi j}). This state contains $N_1=N/2-j$ neutrinos in mass eigenstate $\nu_1$ and $N_2=N/2+j$ neutrinos in mass eigenstate $\nu_2$. The operator $\mathcal{Q}^+(\xi)$ turns one $\nu_2$ into $\nu_1$ such that the state 
\begin{equation} 
\label{BAS 1} 
|\xi\rangle\equiv \mathcal{Q}^+(\xi)|j , -j \rangle_m 
\end{equation} 
contains one more $\nu_1$ and one less $\nu_2$. It is, in fact, a linear superposition of many such states depending on the unknown parameter $\xi$. In order to find for which value(s) of $\xi$ the state in Eq. (\ref{BAS 1}) is an eigenstate, we act on it with the Hamiltonian given in Eq. (\ref{Hamiltonian}). The result is 
\begin{equation} 
\label{BA example} 
{H}\mathcal{Q}^+(\xi)|j , -j \rangle_m=\left( E_{(-j)}-\xi-2\mu j \right)\mathcal{Q}^+(\xi)|j , -j \rangle_m - 
\left(1+2\mu \sum_{\omega}\frac{-j_{\omega}}{\omega-\xi} \right)\mathcal{J}^+|j , -j \rangle_m~. 
\end{equation} 
Here $E_{(-j)}$ is the energy of the state $|j , -j \rangle_m$ given in Eq. (\ref{Zero Energies j}). In order for the state $\mathcal{Q}^+(\xi)|j , -j \rangle_m$ to be an eigenstate of the Hamiltonian, we should choose $\xi$ in such a way that the second term on the right hand side of Eq. (\ref{BA example}) vanishes. This tells us that $\xi$ should obey 
\begin{equation} 
\label{BAE 1} 
\sum_{\omega}\frac{-j_{\omega}}{\omega-\xi}=-\frac{1}{2\mu}~. 
\end{equation} 
Eq. (\ref{BAE 1}) is called a \emph{Bethe ansatz equation} and in general it has several solutions. It is clear from Eq. (\ref{BA example}) that every $\xi$ which satisfies Eq. (\ref{BAE 1}) gives us an eigenstate in the form of Eq. (\ref{BAS 1}) with the energy 
\begin{equation} 
\label{Energy 1} 
E(\xi)=E_{(-j)}-\xi-2\mu j~. 
\end{equation} 
 
One can extend this line of thought to find more eigenstates of the Hamiltonian. In the most general case, the Gaudin operator plays the role of a one parameter step operator, i.e., the eigenstates can be obtained by its repeated application on a lowest weight state but the parameter $\xi$ takes on different values at each step. It can be shown that a state in the form 
\begin{equation} 
\label{BAS +} 
|\xi_1,\xi_2,\dots\xi_\kappa\rangle\equiv \mathcal{Q}^+(\xi_1)\mathcal{Q}^+(\xi_2)\dots\mathcal{Q}^+(\xi_\kappa)|j , -j \rangle_m 
\end{equation} 
is an eigenstate of the Hamiltonian with the energy 
\begin{equation} 
E(\xi_1, \xi_2,\dots,\xi_\kappa)=E_{(-j)}-\sum_{\alpha=1}^\kappa \xi_{\alpha}-\kappa\mu(2j-\kappa+1)~, 
\label{Energy +} 
\end{equation} 
if the Bethe ansatz variables $\xi_1, \xi_2,\dots,\xi_\kappa$ obey the \emph{Bethe ansatz equations} 
\begin{equation} 
\sum_{\omega}\frac{-j_{\omega}}{\omega-\xi_\alpha}=-\frac{1}{2\mu}+\sum_{\substack{\beta=1\\ \left(\beta\neq\alpha\right)}}^{\kappa}\frac{1}{\xi_{\alpha}-\xi_{\beta}}~. 
\label{BAE +} 
\end{equation}  
Eqs. (\ref{BAE +}) form a set of $\kappa$ complex equations in $\kappa$ complex variables that have to be simultaneously satisfied for every $\alpha=1,2,\dots,\kappa$. As stated earlier, the state $|j , -j \rangle_m$ contains respectively $N_1=N/2-j$ and $N_2=N/2+j$ neutrinos in mass eigenstates $\nu_1$ and $\nu_2$. Since each one of the Gaudin operators $\mathcal{Q}^+(\xi_\alpha)$ transform one $\nu_2$ into $\nu_1$, corresponding occupancies for the state in Eq. (\ref{BAS +}) are $N_1=N/2-j+\kappa$ and $N_2=N/2+j-\kappa$. It should be noted that if we add one more Gaudin operator to the state in Eq. (\ref{BAS +}) to transform one more neutrino from $\nu_2$ to $\nu_1$, then the Bethe ansatz equations (\ref{BAE +}) are modified and become a \emph{new} set of $\kappa+1$ equations in $\kappa+1$ variables. This new set of equations are different from the earlier ones with $\kappa$ variables and have different solutions which could be denoted as $(\check{\xi}_1,\check{\xi}_2,\dots,\check{\xi}_\kappa,\check{\xi}_{\kappa+1})$. Each time one adds one more Gaudin operator to Eq. (\ref{BAS +}) to find new eigenstates with different occupancies, one has to solve a new set of coupled algebraic equations. Note that the example described in Eqs. (\ref{BAS 1}-\ref{Energy 1}) is a special case of the general scheme described in Eqs. (\ref{BAS +}-\ref{BAE +}) with $\kappa=1$. In this particular case, the sum on the right hand side of Eq. (\ref{BAE +}) vanishes because there is only one Bethe ansatz variable. 
 
It is obvious from the above remarks that it becomes more and more difficult to find eigenstates as we flip more and more neutrinos from $\nu_2$ into $\nu_1$. However, a symmetry transformation between the eigenstates of the Hamiltonian is helpful in reducing the number of Bethe ansatz equations that has to be solved. In order to present this symmetry, let us first observe that the operator 
\begin{equation} 
{T}=e^{-i\pi{\mathcal{J}}^{x}} 
\end{equation} 
transforms $\nu_{1}$ and $\nu_{2}$ neutrinos into each other. In particular, it exchanges the states $|j , -j \rangle_m$ and $|j , j \rangle_m$ defined in Eqs. (\ref{Define Phi j}) 
\begin{equation} 
{T}^{\dagger}|j , -j \rangle_m=|j , j \rangle_m 
\end{equation} 
and it also transforms the isospin operators as    
\begin{equation} 
{T}^{\dagger}\mathcal{J}_{\omega}^{\pm}{T}=\mathcal{J}_{\omega}^{\mp}\qquad\mbox{and}\qquad {T}^{\dagger}\mathcal{J}_{\omega}^z{T}=-\mathcal{J}_{\omega}^z ~. 
\label{Conversion of Isospin Operators} 
\end{equation} 
As a result, the Hamiltonian given in Eq. (\ref{Hamiltonian in Mass Basis}) is transformed as    
\begin{equation} 
{T}^{\dagger}{H}{T}={H}^\prime=\sum_{\omega}\omega{\mathcal{J}}_{\omega}^z+\vec{\mathcal{J}}\cdot\vec{\mathcal{J}}~. 
\label{Conversion of the Hamiltonian} 
\end{equation} 
Elementary quantum mechanics tells us that if a state $|\psi^\prime\rangle$ is an eigenstate of the Hamiltonian ${H}^\prime={T}^\dagger{H}{T}$ with a particular energy, then the state $|\psi\rangle={T}|\psi^\prime\rangle$ is an eigenstate of the Hamiltonian ${H}$ with the same energy. Therefore, if we find the eigenstates and eigenvalues of the Hamiltonian ${H}^\prime$ with the method presented above (by making the change $\omega \to -\omega$) and then transform the results with the operator ${T}$, we arrive at the eigenstates of the Hamiltonian ${H}$ given in (\ref{Hamiltonian}). In this way, one can show that the states 
\begin{equation} 
\label{BAS -} 
|\xi^\prime_1,\xi^\prime_2,\dots\xi^\prime_\kappa\rangle\equiv \mathcal{Q}^-(\xi^\prime_1)\mathcal{Q}^-(\xi^\prime_2)\dots\mathcal{Q}^-(\xi^\prime_\kappa)|j , j \rangle_m 
\end{equation} 
are eigenstates of the Hamiltonian ${H}$ with the energy 
\begin{equation} 
E(\xi^\prime_1, \xi^\prime_2,\dots,\xi^\prime_\kappa)=E_{(+j)}+\sum_{\alpha=1}^\kappa \xi^\prime_{\alpha}-\kappa\mu(2j-\kappa+1)~, 
\label{Energy -} 
\end{equation} 
if the variables $\xi^\prime_1, \xi^\prime_2,\dots,\xi^\prime_\kappa$ obey the Bethe ansatz equations 
\begin{equation} 
\sum_{\omega}\frac{-j_{\omega}}{\omega-\xi^\prime_\alpha}=\frac{1}{2\mu}+\sum_{\substack{\beta=1\\ \left(\beta\neq\alpha\right)}}^{\kappa}\frac{1}{\xi^\prime_{\alpha}-\xi^\prime_{\beta}}~. 
\label{BAE -} 
\end{equation}  
These Bethe ansatz equations are similar to the ones given in Eq. (\ref{BAE +}) except that the sign of the constant term is reversed. The term 
$E_{(+j)}$ which appears in Eq. (\ref{Energy -}) is the energy of the state $|j , j \rangle_m$ given in Eq. (\ref{Zero Energies j}). The state $|j , j \rangle_m$ which is defined in Eq. (\ref{Define Phi j}) contains respectively $N_1=N/2+j$ and $N_2=N/2-j$ neutrinos in mass eigenstates $\nu_1$ and $\nu_2$. Each Gaudin operator $\mathcal{Q}^-(\xi^\prime_\alpha)$ transforms one $\nu_1$ into $\nu_2$ so that the corresponding occupancies of the state in Eq. (\ref{BAS -}) are $N_1=N/2+j-\kappa$ and $N_2=N/2-j+\kappa$. We see that the occupancies of the mass eigenstates $\nu_1$ and $\nu_2$ are reversed in the state given in Eq. (\ref{BAS -}) as compared to the state given in Eq. (\ref{BAS +}) but both states are reached by solving the same number of equations. Therefore, depending on the occupancies of the mass eigenstates $\nu_1$ and $\nu_2$, either the method presented in Eqs. (\ref{BAS +}-\ref{BAE +}) or the method presented in Eqs. (\ref{BAS -}-\ref{BAE -}) is more economical to use.

\subsection{Electrostatic Analogy} 
\label{Subsection: Electrostatic Analogy} 
 
\begin{figure}[t] 
\begin{center} 
\includegraphics[scale=0.3]{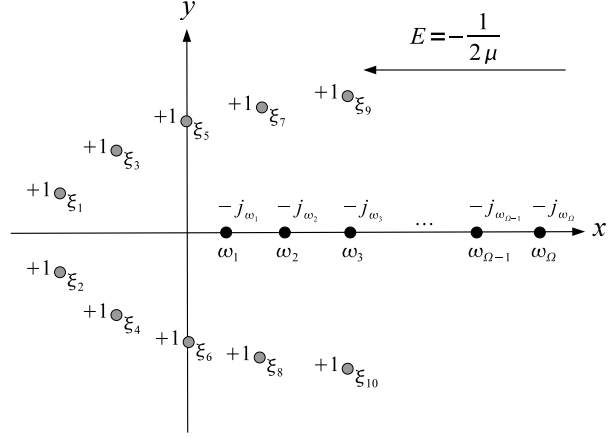} 
\caption{The Bethe ansatz equations (\ref{BAE +}) can be viewed as the stability conditions for a group of point charges in two dimensions as shown here.} 
\label{Fig: Electrostatic_Analogy} 
\end{center} 
\end{figure} 
 
The Bethe ansatz equations given in Eqs. (\ref{BAE +}) and (\ref{BAE -}) can be viewed as the stability conditions for a group of point charges in two dimensions\footnote{Alternatively, one can imagine infinite lines of charge in three dimensions perpendicular to a plane.} \cite{Gaudin Electrostatic Analogy}. Let us consider Eqs. (\ref{BAE +}) for which the electrostatic model is depicted in Fig. \ref{Fig: Electrostatic_Analogy}. There are two kinds of charges in this scheme: the \emph{fixed negative charges} and the \emph{free positive charges}. For each energy mode we have a fixed point charge of magnitude $-j_{\omega}$ pinned at the point $\omega$ on the real axis and for each Bethe ansatz variable we have a free charge of magnitude $+1$ whose equilibrium position yields $\xi_\alpha$. There is also a constant and uniform electric field in the $-x$ direction as shown in the figure.   
 
In two dimensions, one can use a complex coordinate $z=x+iy$ instead of two real coordinates $(x,y)$. The electrostatic potential at the point $z$ created by a charge $q$ sitting at the point $z_0$ is proportional to $q\ln{|z-z_0|}$. As a result, the total electrostatic energy of the charge configuration shown in Fig. \ref{Fig: Electrostatic_Analogy} is proportional to 
\begin{equation} 
\label{Electrostatic Energy} 
V\propto \frac{1}{2\mu}\sum_{\alpha} Re(\xi_\alpha)-\frac{1}{2\mu} \sum_{\omega} j_{\omega} Re(\omega) 
-\frac{1}{2}\sum_{\substack{\alpha,\beta\\ (\alpha\neq\beta)}} \ln{|\xi_\alpha-\xi_\beta|} 
-\frac{1}{2}\sum_{\substack{\omega,\omega^\prime \\ \omega\neq \omega^\prime}} j_{\omega} j_{\omega^\prime}\ln{|\omega-\omega^\prime|}+ 
\sum_{\alpha,\omega} j_{\omega} \ln{|\xi_\alpha-\omega|}~. 
\end{equation} 
The free charges come to an equilibrium when the electrostatic potential energy reaches a local minimum. The minimum energy condition is obtained by setting  
\begin{equation} 
\label{Electrostatic Condition} 
\frac{\partial V}{\partial \xi_\alpha}=0~, 
\end{equation} 
for every $\alpha=1,2\dots,\kappa$. It is a straightforward calculation to show that Eqs. (\ref{Electrostatic Energy}) and (\ref{Electrostatic Condition}) lead to the Bethe ansatz equations (\ref{BAE +}) which now tell that the total force on each free charge is zero. A similar electrostatic picture for the Bethe ansatz equations (\ref{BAE -}) can be obtained by reversing the direction of the electric field in Fig. \ref{Fig: Electrostatic_Analogy}. Note that since the solutions of Bethe ansatz equations come in complex conjugate pairs, the organization of the free charges is always symmetric with respect to $x$ axis.
 
The electrostatic analogy described here provides an intuitive picture and can be particularly useful in finding the solutions of Bethe ansatz equations in the limit of a large number of neutrinos. In this scheme, the total amount of fixed charge on the $x$ axis is equal to $-j$. Although $j$ takes all values between $0$ and $N/2$, for most eigenstates its value is very large. The total amount of free charges is given by 
\begin{equation}
\label{kappa}
\kappa=j+(N_1-N_2)/2~.
\end{equation}
Here we assume that $j$ and $\kappa$ are of same order of magnitude which is true for most eigenstates.
 
Suppose that we start with a small number of neutrinos in a very small volume so that the neutrino density is in the regime where the neutrino-neutrino interactions are important. In this case we have a small amount of charge on the $x$ axis, a few free charges for a typical eigenstate and a small external electric field in the $-x$ direction. Note that, according to the definition of $\mu$ given in Eq. (\ref{Define mu}), the external electric field $-1/2\mu$ is proportional to the volume occupied by the neutrinos. Now, suppose that we increase the number of neutrinos and the volume which they occupy by keeping the density constant. In this case the external electric field grows proportionally while the total fixed charge on the $x$ axis and the number of free charges also increase. For a realistic neutrino spectrum, fixed charges form a continuous charge distribution on the $x$ axis. However, for the sake of this discussion, we assume that they are combined into discreet energy bins such that each $j_{\omega}$ is of the order of $j$. Numerical solutions of Bethe ansatz equations suggest that in this limit the free charges form (piece-wise) continuous distributions. Such solutions have been worked out in the context of the electron pairing in superconductors \cite{Dukelsky:2004re,Richardson:1966zza,Richardson Large N,Gaudin Electrostatic Analogy,Roman:2002dh,Amico:2002hq}. An intuitive way of dealing with this limit was introduced in Ref. \cite{Richardson Large N} and is based on an expansion of the total electrostatic field in powers of $1/\kappa$. In what follows we closely follow this reference. 

Let us begin by considering the total electrostatic field of the system in Fig. \ref{Fig: Electrostatic_Analogy}. It is given by
\begin{equation}
\label{Field}
F(z)=-\frac{1}{2\mu}+\sum_{\omega}\frac{-j_{\omega}}{z-\omega}+\sum_{\alpha=1}^\kappa\frac{1}{z-\xi_\alpha}~.
\end{equation} 
Using the Bethe ansatz equations (\ref{BAE +}), it is straightforward to show that the electrostatic field obeys the following differential equation:
\begin{equation}
\label{Differential Equation}
\frac{dF}{dz}+F^2=\frac{1}{2}\sum_{\omega}\frac{j_{\omega}}{(z-\omega)^2}+\left(\sum_{\omega}\frac{-j_{\omega}}{z-\omega}+\frac{1}{2\mu}\right)^2-\sum_{\omega}\frac{j_{\omega} H(\omega)}{z-\omega}~.
\end{equation}
Here $H(\omega)$ is the electric field produced by the free charges at the position of the fixed charge $-j_\omega$. It is given by
\begin{equation}
\label{H_w}
H(\omega)=\sum_{\alpha=1}^\kappa \frac{1}{\omega-\xi_\alpha}~.
\end{equation}
$H(\omega)$ can also be expressed as a contour integral  
\begin{equation}
\label{H_w Contour}
H(\omega)=\frac{1}{2\pi i}\oint_C\frac{F(z)}{\omega-z}~.
\end{equation}
Here $C$ is a contour which encloses only those singularities of the field $F(z)$ due to the free charges. The equivalence of Eqs. (\ref{H_w}) and (\ref{H_w Contour}) can be shown by direct substitution of the field $F(z)$ given in Eq. (\ref{Field}) in Eq. (\ref{H_w Contour}).

Let us now consider the multipole expansion of the electrostatic field $F(z)$:
\begin{equation}
F(z)=\sum_{m=0}^\infty F^{(m)} z^{-m}~. 
\end{equation}
A direct expansion of Eq. (\ref{Field}) gives these multipole moments as
\begin{eqnarray}
F^{(0)}(z)&=&-\frac{1}{2\mu}~, \label{Zeroth Moment}\\
F^{(1)}(z)&=&\kappa-j~, \label{First Moment}\\
F^{(n)}(z)&=&\sum_{\alpha=1}^\kappa \xi_\alpha^n-\sum_{\omega}\omega^n j_{\omega} \qquad n\geq 2~, \label{Higher Moment}
\end{eqnarray} 
for $|z|>\max(\omega,|\xi_\alpha|)$. The zeroth order moment gives the value of the field at infinity and therefore is equal to the value of the constant field. The first order moment is equal to the total amount of charge, as expected. By comparing Eqs. (\ref{Higher Moment}) and (\ref{Energy +}) one can see that the energy of an eigenstate can be found from the second order moment of the corresponding electrostatic field. 

Richardson's method is based on solving the differential equation given in Eq. (\ref{Differential Equation}) for the leading terms in an expansion of the form 
\begin{equation}
\label{Expansion}
F(z)=\sum_{r=0}^\infty F_r(z)~. 
\end{equation} 
Here the term $F_r(z)$ is assumed to be of the order of $\kappa^{1-r}$. In the present case, we will consider only the $r=0$ term which is of the order of $\kappa$. It is reasonable to assume that those terms on the right hand side of the differential equation which are of the form $\sum_{\omega} j_{\omega} (\dots)$ are of the order of $j$. Since the volume is increased in proportion with the number of neutrinos, we also treat the external field $-1/2\mu$ as being of the order of $j$. Finally, Eq. (\ref{H_w}) tells us that $H_{\omega}$ is of the order of $\kappa$. Taking these into account and substituting the expansion given in Eq. (\ref{Expansion}) into the differential equation (\ref{Differential Equation}) we find
\begin{equation}
\label{F_0 Equation}
F_0(z)^2=\left(\sum_{\omega}\frac{j_{\omega}}{\left(z-\omega\right)^2}+\frac{1}{2\mu}\right)^2-\sum_{\omega}\frac{j_{\omega}H_0(\omega)}{z-\omega}~, 
\end{equation}   
where $H_0(\omega)$ is given by 
\begin{equation}
H_0(\omega)=\frac{1}{2\pi i}\oint_C\frac{F_0(z)}{\omega-z}~. 
\end{equation}
Eq. (\ref{F_0 Equation}) is an integral equation which includes only the fixed charges as parameters and should be solved to find the leading order electrostatic field created by the whole configuration including the free charges in their stable configuration. For this reason, it can be thought as a replacement for Bethe ansatz equations. Once a solution of Eq. (\ref{F_0 Equation}) is obtained, one can determine the leading order contribution to the energy of the corresponding eigenstate from the second order moment of the field. The locations of the free charges can be found from the singularities (or the branch cuts) of the field other than the fixed charges and these can be used to write down the corresponding eigenstate itself.  

However, instead of directly solving Eq. (\ref{F_0 Equation}), a more practical way is to form its solutions on physical grounds. The typical way to proceed is to start with a small number of neutrinos in a small volume and see the way the solutions of Bethe ansatz equations organize themselves as the number of neutrinos are increased while the density is kept at a constant value. As mentioned before, numerical simulations suggest that the free charges coalesce to form (piece-wise) continuous charge distributions. One can guess the limiting shape that they will assume possibly in terms of some unknown parameters. These parameters can later be determined either by forming a field $F_0(z)$ which describes such a distribution to the leading order and substituting it in Eq. (\ref{F_0 Equation}) or, in simple situations, directly from self consistency requirements as described in the following example.

For example, it is known that a solution of Bethe ansatz equations exists in which all free charges organize themselves into a single arc extending from a point $a$ to a point $a^*$ (the parameters to be determined). The situation is similar to the one depicted in Fig. \ref{Fig: Electrostatic_Analogy} except that the free charges now form a continuous curve. In the case of the BCS model, this solution is known to lead to the BCS ground state. Since the singularities due to the free charges form a continuum, the field $F_0(z)$ is expected to have a branch cut along the arc such that
\begin{equation}
\frac{1}{2\pi i}\oint_{C}F_0(z)dz=\kappa
\end{equation}
is satisfied for any closed path $C$ enclosing the free charges. Note that it is reasonable to also demand that the field $F_0(z)$ describes the fixed charges and the external field correctly, i.e.,
\begin{equation}
\lim_{z\to\omega}(z-\omega)F_0(z)=j_{\omega} \qquad\mbox{and}\qquad
\lim_{z\to\infty}F_0(z)=-\frac{1}{2\mu}
\end{equation}
because both $j_{\omega}$ and the external field are of the order of $j$. In this case, the zeroth and the first order moments given in  Eqs. (\ref{Zeroth Moment}) and (\ref{First Moment}) will receive no corrections from the higher order terms in the expansion of electrostatic field given in Eq. (\ref{Expansion}).  

In Ref. \cite{Richardson Large N}, it is argued that a candidate for the field $F_0(z)$ with these properties is given by
\begin{equation}
\label{F_0 Solution}
F_0(z)=-[(z-a)(z-a^*)]^{1/2}\sum_{\omega}\frac{j_{\omega}}{|\omega-a|(z-\omega)}~. 
\end{equation}
This field is of the order of $j$ and has a branch cut along an arc which extends from $a$ to $a^*$. It also has singularities at the points $\omega$ and a constant limit as $z\to\infty$ as required by the above arguments. The first three moments of the field $F_0(z)$ are given by
\begin{eqnarray}
F_0^{(0)}&=&-\sum_{\omega}\frac{j_{\omega}}{|\omega-a|}~,\label{Zeroth Moment of F_0}\\
F_0^{(1)}&=&\sum_{\omega}\frac{j_{\omega}(\lambda-\omega)}{|\omega-a|}~,\label{First Moment of F_0}\\
F_0^{(2)}&=&\sum_{\omega}\frac{j_{\omega}(\lambda\omega-\omega^2-\Delta^2/2)}{|\omega-a|}~,\label{Second Moment of F_0}
\end{eqnarray}
where $a=\lambda+i\Delta$. One can try to determine the value of the parameter $a$ by demanding that $F_0(z)$ reproduces the zeroth and the first order moments without any contributions from higher order fields as mentioned above. Equating the moments in Eqs. (\ref{Zeroth Moment of F_0}-\ref{First Moment of F_0}) to those in Eqs. (\ref{Zeroth Moment}-\ref{First Moment}) one finds the equations
\begin{equation}
\frac{1}{2\mu}=\sum_{\omega}\frac{j_{\omega}}{|\omega-a|}
\qquad\qquad
\frac{\lambda}{2\mu}+j-\kappa=\sum_{\omega} \frac{\omega j_{\omega}}{|\omega-a|}\label{BCS Equations}
\end{equation} 
whose solution yields the value of $a=\lambda+i\Delta$. It is shown in Ref. \cite{Richardson Large N} that the substitution of $F_0(z)$ given in Eq. (\ref{F_0 Solution}) with the value of $a$ satisfying Eqs. (\ref{BCS Equations}) shows that it is indeed the correct solution of the differential equation (\ref{F_0 Equation}). Note that, in the context of the BCS theory, the parameters $\Delta$ and $\lambda$ are known as the BCS gap and the chemical potential, respectively. Accordingly, Eqs. (\ref{BCS Equations}) are called the BCS gap and chemical potential equations.

It was already mentioned that the second order moment of the total electrostatic field is related to the energy of the corresponding eigenstate. Using Eqs. (\ref{Energy +}), (\ref{Higher Moment}), (\ref{Second Moment of F_0}) and (\ref{BCS Equations}), the energy of the eigenstate corresponding to the solution given in Eq. (\ref{F_0 Solution}) can be found as 
\begin{equation}
E=E_{(-j)}-\kappa\mu(2j-\kappa+1)-\sum_{\omega} j_{\omega}\omega\left(1-\frac{\omega-\lambda}{|\omega-a|}\right)+\frac{\Delta^2}{4\mu}
\end{equation} 
to the first order in $j$. Here $E_{(-j)}$ is given by Eq. (\ref{Zero Energies j}). 

Note that the neutrino Hamiltonian given in Eq. (\ref{Hamiltonian in Mass Basis}) has an overall minus sign with respect to the BCS pairing Hamiltonian given in Eq. (\ref{BCS Hamiltonian}). Therefore, although the field $F_0(z)$ given in Eq. (\ref{F_0 Solution}) yields the ground state of the (reduced) BCS pairing Hamiltonian, for the neutrino case, it yields the eigenstate with the highest energy. A study of the general properties of Bethe ansatz states in the continuum limit from the perspective of collective neutrino oscillations is an open problem and will be considered elsewhere.  
\section{The Random Phase Approximation (RPA)} 
\label{Section: The Random Phase Approximation} 
 
A large quantum system such as a self interacting neutrino gas can be conveniently studied within an effective one-particle approximation whereby it is described in terms of single particles interacting with an average potential created by all other particles in the medium. The requirement that the average potential should evolve in line with the time evolution of individual particles gives rise to a set of consistency equations. In this section, we formulate the effective one-particle description in terms of the RPA method. Our aim is to reach a quantum mechanical effective one-body Hamiltonian which describes the interaction of a single neutrino with its background represented by the classical polarization vectors. In this picture, the well known time evolution equations of the polarization vectors emerge from the RPA consistency condition. In Section \ref{Subsection: Equations of Motion}, we show that the time evolution equations of the polarization vectors possess several constants of motion which are basically the expectation values of the many-body constants of motion mentioned in Section \ref{Subsection: The Hamiltonian and the Quantum Invariants}. One of the main results of Section \ref{Section: The Random Phase Approximation} involves the diagonalization of the linearized RPA Hamiltonian which yields a noninteracting (quasi-particle) basis for neutrinos. We show in Section \ref{Subsection: Stationary Solutions and the Spectral Splits} that the well known phenomenon of spectral splits can be viewed as a result of the adiabatic evolution of these quasi-particle degrees of freedom from flavor to mass eigenstates under appropriate conditions.    
\subsection{Equations of Motion} 
\label{Subsection: Equations of Motion} 

In the method of RPA, a quadratic operator ${\cal O}_1{\cal O}_2$ is approximated as 
\begin{equation} 
\label{RPA approximation} 
{\cal O}_1 {\cal O}_2 \sim  
{\cal O}_1 \langle {\cal O}_2 \rangle + \langle {\cal O}_1 \rangle {\cal O}_2 - 
\langle {\cal O}_1 \rangle \langle {\cal O}_2 
\rangle~, 
\end{equation} 
where the expectation values should be calculated with respect to a state $|\Psi\rangle$ which satisfies the condition 
\begin{equation} 
\label{RPA condition} 
\langle {\cal O}_1  {\cal O}_2 \rangle = \langle {\cal O}_1 \rangle \langle {\cal O}_2\rangle~. 
\end{equation} 
The state $|\Psi\rangle$ is usually found by solving the resulting \emph{RPA consistency equations} as illustrated below. 
 
Application of the RPA method to the neutrino Hamiltonian given in Eq. (\ref{Hamiltonian}) yields  
\begin{equation} 
\label{RPA Hamiltonian} 
{H}\sim{H}^{\mbox{\tiny RPA}}=\sum_{\omega}\omega\vec{B}\cdot\vec{J}_{\omega} 
+\mu\vec{P}\cdot\vec{J}~. 
\end{equation} 
In Eq. (\ref{RPA Hamiltonian}), we introduced the \emph{polarization vector} which is defined as 
\begin{equation}  
\vec{P}_{\mathbf{p}}=2\langle\vec{J}_{\mathbf{p}}\rangle~. 
\label{Polarization Vectors} 
\end{equation} 
The total polarization vectors are defined as in Eq. (\ref{Totals}). In Eq. (\ref{Polarization Vectors}), the expectation values are calculated with respect to a state $|\Psi\rangle$ which is assumed to be a tensor product of one-particle states, i.e.,
\begin{equation} 
\label{RPA State} 
|\Psi\rangle\equiv |\psi(\mathbf{p}_1)\rangle\otimes |\psi(\mathbf{p}_2)\rangle \otimes\dots\otimes|\psi(\mathbf{p}_N)\rangle~, 
\end{equation} 
where 
\begin{equation} 
|\psi(\mathbf{p})\rangle=\psi_e (\mathbf{p}) |\nu_e\rangle + \psi_x(\mathbf{p}) |\nu_x\rangle~. 
\end{equation} 
The state $|\Psi\rangle$ defined in Eq. (\ref{RPA State}) is an $SU(2)$ coherent state and thus automatically obeys the condition given in Eq. (\ref{RPA condition}) (see Ref. \cite{Balantekin:2006tg}). Substitution of the state $|\Psi\rangle$ in Eq. (\ref{Polarization Vectors}) yields the explicit form of the polarization vector as follows\footnote{\label{Footnote: Neutrino Polarization Vectors}An alternative way to define the neutrino polarization vector is to use the density matrix method. For a single neutrino with momentum $\mathbf{p}$, the elements of the flavor density matrix are defined as $\rho_{\alpha\beta}(\mathbf{p})=\langle a_\beta^\dagger(\mathbf{p}) a_\alpha(\mathbf{p})\rangle$. For the state $|\Psi\rangle$ given in Eq. (\ref{RPA State}), this definition amounts to 
\begin{equation*} 
\rho(\mathbf{p})= 
\left( 
\begin{array}{cc} 
|\psi_e(\mathbf{p})|^2  &  \psi_x^*(\mathbf{p}) \psi_e(\mathbf{p})   \\ 
\psi_e^*(\mathbf{p}) \psi_x(\mathbf{p})  & |\psi_x(\mathbf{p})|^2   
\end{array} 
\right)~. 
\end{equation*}  
Then, one can define the polarization vector though the expansion $\rho(\mathbf{p})=\frac{1}{2}\left(1+\vec{\sigma}\cdot\vec{P}_\mathbf{p}\right)$ in terms of the Pauli sigma matrices $\vec{\sigma}$. This definition leads to the same result as in Eq. (\ref{Polarization Vectors Explicit}). 
}: 
\begin{equation} 
\label{Polarization Vectors Explicit} 
\vec{P}_{\mathbf{p}}= 
\left( 
\begin{array}{c} 
2 Re[\psi_e^*(\mathbf{p}) \psi_x(\mathbf{p})]   \\ 
2 Im[\psi_e^*(\mathbf{p}) \psi_x(\mathbf{p})]  \\ 
|\psi_e(\mathbf{p})|^2-|\psi_x(\mathbf{p})|^2   
\end{array} 
\right)~. 
\end{equation}  
The linearized Hamiltonian given in Eq. (\ref{RPA Hamiltonian}) yields the following Heisenberg equations of motion 
\begin{equation} 
\label{RPA EoM for J} 
\frac{d}{dt}\vec{J}_{\omega}=-i[\vec{J}_{\omega},{H}^{\mbox{\tiny RPA}}]=(\omega\vec{B}+\mu\vec{P})\times\vec{J}_{\omega}~. 
\end{equation} 
Due to the self consistency requirement of the RPA method, the polarization vectors should then obey 
\begin{equation} 
\label{RPA EoM} 
\frac{d}{dt}\vec{P}_{\omega}=(\omega\vec{B}+\mu\vec{P})\times\vec{P}_{\omega} 
\end{equation} 
for every $\omega$. Eq. (\ref{RPA EoM}) can be obtained by taking the expectation values of the isospin operators on both sides of Eq. (\ref{RPA EoM for J}). Solution of the consistency equations yields the state $|\Psi\rangle$ postulated at Eq. (\ref{RPA State}).  
 
Note that in approximating the exact many-body Hamiltonian with the RPA method, one sacrifices the conservation of the many-body invariants ${h}_{\omega}$ given in Eq. (\ref{Invariants}) because the RPA Hamiltonian no longer commutes with these operators. However, the expectation values of the operators ${h}_{\omega}$ with respect to the state $|\Psi\rangle$ continue to be invariant. The expectation values $\langle{h}_{\omega}\rangle$ can be easily calculated using Eqs. (\ref{RPA condition}) and (\ref{Polarization Vectors}). The result is  
\begin{equation} 
\label{Invariants RPA} 
I_{\omega}=2\langle {h}_{\omega} \rangle=\vec{B}\cdot\vec{P}_{\omega}+\mu\sum_{\omega^\prime\left(\neq\omega\right)}\frac{\vec{P}_{\omega}\cdot\vec{P}_{\omega^\prime}}{\omega-{\omega^\prime}}~. 
\end{equation} 
Here a factor of $2$ is introduced in the definition of $I_{\omega}$ for convenience. Using the RPA equations of motion (\ref{RPA EoM}), it is a straightforward calculation to show that 
\begin{equation} 
\label{Conservation RPA} 
\frac{d}{dt} I_{\omega} =0~. 
\end{equation} 
The particular combination  
\begin{equation}
\label{Conservation of K0} 
K_0\equiv\sum_{\omega} I_{\omega} =\vec{B}\cdot\vec{P} 
\end{equation} 
is already pointed out to be a constant of motion by many authors. Investigation of the N-mode coherence in collective neutrino oscillations leads to similar invariants obtained as the scalar product of the polarization vector and its Hilbert transform \cite{Raffelt:2011yb}. The invariants we find {\it in the RPA formalism} are closely related to  the invariants of such N-mode coherence. It should be mentioned out that such invariants were first discussed in the context  
of BCS dynamics \cite{yuzb}. 
 
In the mass basis, we denote the polarization vectors by calligraphic letters as in the case of isospin operators, i.e., $\vec{\mathcal{P}}_{\mathbf{p}}=2\langle\vec{\mathcal{J}}_{\mathbf{p}}\rangle$. From Eq. (\ref{Mass Isospin Operators}) we see that the conserved quantity given in Eq. (\ref{Conservation of K0}) is equal to
\begin{equation}
\label{K0} 
K_0=-\mathcal{P}^z=\frac{\langle\hat{N}_2\rangle-\langle\hat{N}_1\rangle}{2}~. 
\end{equation} 
Together with the conservation of $\langle\hat{N}_1\rangle+\langle\hat{N}_2\rangle$, this implies that both $\langle\hat{N}_1\rangle$ and $\langle\hat{N}_2\rangle$ are individually conserved under the RPA evolution. Therefore, although the exact many-body Hamiltonian given in Eq. (\ref{Hamiltonian}) strictly conserves the number of particles in each mass eigenstate (see Eq. (\ref{Particle Number Conservation})), the RPA evolution conserves only their average values.   
 
 
\subsection{Stationary Solutions and the Spectral Splits} 
\label{Subsection: Stationary Solutions and the Spectral Splits} 
 
It has been pointed out in Section \ref{Subsection: The Hamiltonian and the Quantum Invariants} that the BCS pairing Hamiltonian and the collective neutrino oscillation Hamiltonian have the same mathematical structure. This equivalence has been exploited in Section \ref{Section: Diagonalization of the Hamiltonian} in order to write down the exact eigenstates and eigenvalues of the many-body neutrino Hamiltonian. In this section, we will consider the eigenstates and eigenvalues of the RPA neutrino Hamiltonian given in Eq.  (\ref{RPA Hamiltonian}). These eigenstates correspond to the steady state solutions of Eqs. (\ref{RPA EoM}). To this end, we will borrow the method of Bogoliubov transformations from the RPA-BCS model by which the well known BCS ground state and its excitations are obtained.  
 
Application of the method of RPA to the BCS pairing Hamiltonian given in Eq. (\ref{BCS Hamiltonian}) yields 
\begin{equation} 
{H}_{\mbox{\tiny BCS}}^{\mbox{\tiny RPA}}= 
\sum_{k}2{\epsilon_{k}}{t}_{k}^z-G\left(\langle{T}^+\rangle{T}^- + {T}^+\langle{T}^-\rangle\right)~. 
\label{RPA BCS Hamiltonian} 
\end{equation} 
In order to find the ground state of the RPA-BCS Hamiltonian for a given number of pairs, one introduces a chemical potential through the method of Lagrange variables. In other words, one tries to minimize the Hamiltonian  
\begin{equation} 
{H}_{\mbox{\tiny BCS}}^{\mbox{\tiny RPA}}-2 \mu {T}^z= 
\sum_{k}2({\epsilon_{k}-\mu}){t}_{k}^z-G\left(\langle{T}^+\rangle{T}^- + {T}^+\langle{T}^-\rangle\right)~. 
\label{BCS Hamiltonian with chemical potential} 
\end{equation} 
The excited states are then obtained from the ground state by applying the excitation operators. Although the many-body BCS Hamiltonian ${H}_{\mbox{\tiny BCS}}$ conserves the total pair number $2{T}^z$ (see Eq. (\ref{Conservation of T3})), the RPA-BCS Hamiltonian given above conserves only its mean value $2\langle{T}^z\rangle$. Thus, the BCS ground state and the excited states obtained this way have a constant mean pair number, but they are not eigenstates of the pair number operator, i.e., they have \emph{indefinite} number of pairs.

As pointed out at the end of Section \ref{Subsection: Equations of Motion}, one has a similar situation for neutrino Hamiltonian and we introduce an analogous Lagrange variable which we denote by $\omega_c$: 
\begin{equation} 
{H}^{\mbox{\tiny RPA}}+\omega_c{\mathcal{J}}^z=\sum_{\omega}(\omega_c-\omega){\mathcal{J}}_{\omega}^z+\mu\vec{\mathcal{P}}\cdot\vec{\mathcal{J}}~. 
\label{RPA Hamiltonian with w_c} 
\end{equation}
This way one can find the state which minimizes the RPA neutrino Hamiltonian (\ref{RPA Hamiltonian}) for a given value of the constant in Eq. (\ref{K0}). Note that, since the Hamiltonian in Eq. (\ref{RPA Hamiltonian with w_c}) commutes with the total particle number operator $\hat{N}$, the total neutrino number $N$ is always well defined. Consequently, the Lagrange multiplier in Eq. (\ref{RPA Hamiltonian with w_c}) fixes the values of both $\langle\hat{N}_1\rangle$ and $\langle\hat{N}_1\rangle$.  
In what follows, we will use a method borrowed from Bogoliubov's solution of the RPA-BCS model which yields both the ground and the excited states with these properties \cite{Bogoliubov}.  

Let us begin by pointing out that the Hamiltonian in Eq. (\ref{RPA Hamiltonian with w_c}) can be diagonalized by rotating the flavor space of each energy mode in a different way as described in Section \ref{Subsection: Mass and Flavor Isospin Operators}. Such a rotation can be carried out by the unitary operator given in Eq. (\ref{General Unitary Operator}). This time we denote the operator by $U_{\bbomega}$ and its parameter by $z_{\omega}$ to emphasize that all neutrinos with a given energy are subject to the same transformation while different energy modes undergo different transformations. Our aim is to write the Hamiltonian given in Eq. (\ref{RPA Hamiltonian with w_c}) in the form 
\begin{equation} 
\label{RPA Hamiltonian with U'} 
{H}^{\mbox{\tiny RPA}}+\omega_c{\mathcal{J}}^z=\sum_{\omega}   
2\lambda_{\omega}  {U_{\bbomega}^\dagger} {\mathcal{J}}^z_{\omega} {U_{\bbomega}}~. 
\end{equation} 
The first of Eqs. (\ref{General Transformation}) tells us that for the equality in Eq. (\ref{RPA Hamiltonian with U'}) to hold, the rotation parameter $z_{\omega}=e^{i\delta_{\omega}}\tan\theta_{\omega}$ should satisfy 
\begin{eqnarray}
2\lambda_{\omega}\cos2\theta_{\omega}&=&(\omega_c-\omega)+\mu\mathcal{P}^z~, \nonumber \\
2\lambda_{\omega}e^{\pm i\delta_{\omega}}\sin2\theta_{\omega}&=&\mu\mathcal{P}^\pm~. 
\label{Rotation Parameters}
\end{eqnarray}
Solving Eqs. (\ref{Rotation Parameters}) we find that the phase $\delta_{\omega}$ is the same for all energy modes and is given by
\begin{equation} 
\label{delta} 
e^{i\delta}=\frac{\mathcal{P}^+}{|\mathcal{P}^+|}~ 
\end{equation} 
whereas the rotation angle $\theta_{\omega}$ depends on the energy and is given by  
\begin{equation} 
\label{angles} 
\cos{\theta_{\omega}} =\sqrt{\frac{1}{2}\left(1+\frac{\omega_c-\omega+\mu\mathcal{P}^z}{2\lambda_{\omega}}\right)}~, \qquad 
\sin{\theta_{\omega}} =\sqrt{\frac{1}{2}\left(1-\frac{\omega_c-\omega+\mu\mathcal{P}^z}{2\lambda_{\omega}}\right)}~.  
\end{equation} 
The parameter $\lambda_{\omega}$ can also be found from Eqs. (\ref{Rotation Parameters}) as 
\begin{equation} 
\label{lambda_w} 
\lambda_{\omega}=\frac{1}{2}\sqrt{(\omega_c-\omega+\mu\mathcal{P}^z)^2+\mu^2\mathcal{P}^+\mathcal{P}^-}~. 
\end{equation} 

Eq. (\ref{RPA Hamiltonian with U'}) motivates the definition of the Bogoliubov transformed \emph{quasi-particle} operators\footnote{In its broadest sense, a Bogoliubov transformation is any mixture of particle operators which preserves the (anti) commutation relations. However, the term usually suggests the mixing of creation and annihilation operators because in Bogoliubov's approach to BCS theory the electron and hole states are mixed. Although Eq. (\ref{Bogoliubov Transformation}) is only a rotation in flavor space (apart from the phases) it can also be viewed as a special Bogoliubov transformation in the broad sense of the term because the rotated particle operators satisfy the same anti-commutation relations as the original ones. In that sense, the distinction between the \emph{real} particles and the \emph{quasi}-particles seems somewhat arbitrary because one can see the flavor basis as quasi-particles of the mass basis. However the term quasi-particle is usually reserved for the noninteracting basis in the presence of an interaction between the particles and the background. Here, our usage of the terms \emph{Bogoliubov transformation} and \emph{quasi-particle} is due to the established terminology of the BCS theory.}
\begin{eqnarray} 
\label{Bogoliubov Transformation} 
\alpha_1(\mathbf{p})&=&{U_{\bbomega}^\dagger} a_1(\mathbf{p}){U_{\bbomega}}  
= \cos{\theta_{\omega}} \; a_1(\mathbf{p}) +e^{i\delta}\sin{\theta_{\omega}} \; a_2(\mathbf{p}) \\ 
\alpha_2(\mathbf{p})&=&{U_{\bbomega}^\dagger} a_2(\mathbf{p}){U_{\bbomega}}  
= - e^{-i\delta}\sin{\theta_{\omega}} \; a_1(\mathbf{p})+\cos{\theta_{\omega}} \; a_2(\mathbf{p})\nonumber 
\end{eqnarray} 
together with the corresponding isospin 
\begin{equation} 
\label{alpha Isospin} 
\vec{\mathsf{J}}_{\mathbf{p}}={U_{\bbomega}^\dagger} \vec{\mathcal{J}}_{\mathbf{p}} {U_{\bbomega}}~, 
\end{equation} 
which we call the \emph{$\alpha$-isospin operators}. They are equal to  
\begin{equation} 
{\mathsf{J}}_{\mathbf{p}}^{+}= \alpha_{1}^{\dagger}(\mathbf{p})\alpha_{2}(\mathbf{p})~,\qquad 
{\mathsf{J}}_{\mathbf{p}}^{-}= \alpha_{2}^{\dagger}(\mathbf{p})\alpha_{1}(\mathbf{p})~,\qquad 
{\mathsf{J}}_{\mathbf{p}}^z=\frac{1}{2}\left(\alpha_{1}^{\dagger}(\mathbf{p})\alpha_{1}(\mathbf{p})-\alpha_{2}^{\dagger}(\mathbf{p})\alpha_{2}(\mathbf{p})\right)~. 
\label{alpha Isospin Operators} 
\end{equation} 
The Hamiltonian given in Eq. (\ref{RPA Hamiltonian with U'}) can now be written as the Hamiltonian of a noninteracting system in terms of the quasi-particles 
\begin{equation} 
\label{RPA Hamiltonian in b's} 
{H}^{\mbox{\tiny RPA}}+\omega_c{\mathcal{J}}^z=\sum_{\omega}  
2\lambda_{\omega}  \mathsf{J}^z_{\omega}=\sum_{\mathbf{p}}   
\lambda_{\omega} \left( \alpha_1^\dagger(\mathbf{p})\alpha_1(\mathbf{p})-\alpha_2^\dagger(\mathbf{p})\alpha_2(\mathbf{p})   \right)~. 
\end{equation} 
The eigenstates of the Hamiltonian given in Eq. (\ref{RPA Hamiltonian in b's}) are the ones in which all neutrinos occupy $\alpha_1$ and $\alpha_2$ states. In particular, the ground state is an ``$\alpha_2$ condensate.'' In the language of $\alpha$-isospins introduced in Eq. (\ref{alpha Isospin Operators}), all $\alpha$-isospins point down in the ground state so that it is equal to the lowest weight state 
\begin{equation} 
\label{RPA Ground State} 
|j_{\mbox{\tiny max}},-j_{\mbox{\tiny max}}\rangle_\alpha= \prod_{\mathbf{p}}\alpha_{2}^{\dagger}(\mathbf{p})\;|0\rangle~, 
\end{equation}  
where $j_{\mbox{\tiny max}}=N/2$. The states with higher energies can be obtained by exciting the particles from $\alpha_2$ to $\alpha_1$ states which amounts to applying the operator ${\mathsf{J}}_{\mathbf{p}}^{+}$ defined in Eq. (\ref{alpha Isospin Operators}). In general, the eigenstates and eigenvalues of the Hamiltonian in Eq. (\ref{RPA Hamiltonian in b's}) can be written as  
\begin{equation} 
\label{RPA Eigenstates} 
|\psi\rangle=\prod_{\omega} |j_{\omega} , m_{\omega}\rangle_\alpha \qquad \mbox{and} \qquad E=\sum_{\omega} 2m_{\omega}\lambda_{\omega}~. 
\end{equation} 
In Eqs. (\ref{RPA Ground State}) and (\ref{RPA Eigenstates}), the subscript $\alpha$ points to the fact that these are the eigenstates  of the $\alpha$-isospin\footnote{It is evident from Eq. (\ref{alpha Isospin}) that these states are related to the corresponding eigenstates $|j_{\omega} , m_{\omega}\rangle_m$ of the mass isospin by  
\begin{equation} 
\nonumber 
|j_{\omega} , m_{\omega}\rangle_\alpha={U_{\bbomega}^\dagger} |j_{\omega} , m_{\omega}\rangle_m~, 
\end{equation} 
which in turn are related to the eigenstates $|j_{\omega} , m_{\omega}\rangle_f$  of the flavor isospin by  
\begin{equation} 
\nonumber 
|j_{\omega} , m_{\omega}\rangle_m={U}^\dagger |j_{\omega} , m_{\omega}\rangle_f~. 
\end{equation} 
Therefore, we can write the eigenstates in Eq. (\ref{RPA Eigenstates}) as 
\begin{equation} 
\label{RPA Eigenstates in flavor} 
\nonumber 
|\psi\rangle=\prod_{\omega} |j_{\omega} , m_{\omega}\rangle_\alpha={U_{\bbomega}^\dagger} \prod_{\omega} |j_{\omega} , m_{\omega}\rangle_m={U_{\bbomega}^\dagger} {U}^\dagger \prod_{\omega} |j_{\omega} , m_{\omega}\rangle_f~. 
\end{equation} 
Note that the transformation described by the operator $U_{\bbomega}$ is not a global rotation. From Eq. (\ref{angles}), it is clear that each energy mode $\omega$ undergoes a different rotation. As a result, this transformation does not preserve all scalar products. In particular, the total $\alpha$-isospin quantum number of a state is not always the same as its total mass or flavor isospin quantum number. The state in Eq. (\ref{RPA Ground State}), for example, does not live in the $j_{\mbox{\tiny max}}$ representation of mass or flavor isospin although it lives in the $j_{\mbox{\tiny max}}$ representation of the $\alpha$-isospin. However, $j_\omega$ is a well defined quantum number in all three bases because, within each energy mode $\omega$, the transformation induced by $U_\bbomega$ is a ``global'' one.}. The ground state given in Eq. (\ref{RPA Ground State}) is the special case of Eq. (\ref{RPA Eigenstates}) in which $m_{\omega}=-j_{\omega}$ for all energy modes.  
 
For consistency, the states in Eq. (\ref{RPA Eigenstates}) should satisfy 
\begin{equation} 
\label{Consistency} 
2\langle \psi|{\mathcal{J}}^+|\psi \rangle = \mathcal{P}^+ 
\qquad\mbox{and}\qquad 
2\langle \psi|{\mathcal{J}}^z|\psi \rangle = \mathcal{P}^z~. 
\end{equation} 
It is an easy exercise in algebra to show that  
\begin{equation} 
\label{Mean Value for +} 
2\langle \psi|{\mathcal{J}}^+|\psi \rangle=4\sum_{\omega} m_{\omega} \frac{z_{\omega}}{1+|z_{\omega}|^2} 
=e^{i\delta}\mu|\mathcal{P}^+|\sum_{\omega} \frac{m_{\omega}}{\lambda_{\omega}} 
\end{equation} 
and  
\begin{equation} 
\label{Mean Value for 0} 
2\langle \psi|{\mathcal{J}}^z|\psi \rangle=2\sum_{\omega} m_{\omega} \frac{1-|z_{\omega}|^2}{1+|z_{\omega}|^2} 
=\sum_{\omega} m_{\omega} \frac{\omega-\omega_c+\mu\mathcal{P}^z}{\lambda_{\omega}}~. 
\end{equation} 
In this case, the consistency equations (\ref{Consistency}) become 
\begin{equation} 
\label{Consistency Conditions} 
\frac{1}{2\mu}=\sum_{\omega}\frac{m_{\omega}}{2\lambda_{\omega}} 
\qquad\qquad 
\frac{\omega_c}{2\mu}=\sum_{\omega} \frac{m_{\omega}\omega}{2\lambda_{\omega}}~. 
\end{equation} 
The consistency equations given in Eqs. (\ref{RPA EoM}) differ from those given above in that the former ones are valid for a generic time dependent state whereas the later ones are valid for the eigenstates (i.e., stationary states) of the RPA Hamiltonian. At this point, we would like to note the similarity between Eqs. (\ref{BCS Equations}) and Eqs. (\ref{Consistency Conditions}). From Eq. (\ref{RPA Eigenstates}) we see that the eigenstate with the highest energy has $m_{\omega}=j_{\omega}$. For this particular state, the consistency equations (\ref{Consistency Conditions}) are the same as those given in Eqs. (\ref{BCS Equations}) with the substitution $\lambda=\omega_c-\mu \mathcal{P}^z$ and $\Delta=\mu|\mathcal{P}^+|$ together with $\kappa=j+\mathcal{P}^z/2$ (see Eq. (\ref{kappa})).  
 
A special but very interesting case of Eqs. (\ref{Consistency Conditions}) is obtained in Ref. \cite{Raffelt:2007cb} in connection with the spectral splits of neutrinos which occur if the parameter $\mu$ defined in Eq. (\ref{Define mu}) changes in an adiabatic way from $\infty$ to $0$ (also see Ref. \cite{Duan:2010bg} for an in-depth review). In this case, the Lagrange multiplier $\omega_c$ introduced above plays the role of the split frequency. Here we will show that this phenomenon can also be understood in terms of the adiabatic evolution of quasi-particle states. Namely, we will show that when the neutrino density is high, the quasiparticle states coincide with flavor eigenstates whereas in vacuum they coincide with mass eigenstates. As a result, neutrinos emerging in flavor eigenstates in the high density region gradually transform into mass eigenstates if the neutrino density decreases in an adiabatic way. The Lagrange multiplier $\omega_c$ determines the final energy distribution of mass eigenstates.  
 
In order to see how this phenomenon takes place, let us assume that initially the neutrino density is so high that the $\mu\to\infty$  limit is practically attained and that all neutrinos initially occupy flavor eigenstates $\nu_e$ and $\nu_x$. Note that here it is not assumed that all neutrinos occupy the same flavor state. Neutrinos emanating from the surface of a proto-neutron star after a supernova explosion are believed to meet these initial state assumptions. In $\mu\to\infty$ limit, Eq. (\ref{lambda_w}) gives $2\lambda_{\omega}=\mu|\vec{\mathcal{P}}|=\mu|\vec{P}|$ for all neutrinos. Since we assumed that all neutrinos are in flavor eigenstates, we have $P^+=P^-=0$ in which case Eqs. (\ref{Transformation}) lead to $\mathcal{P}^z=\cos{2\theta}P^z$. Substituting these in Eq. (\ref{angles}) we find 
\begin{equation} 
\label{Angles in the Infinity Limit} 
\cos\theta_{\omega}=\sqrt{\frac{1}{2}\left(1+\frac{P^z}{|\vec{P}|}\cos{2\theta}\right)} \qquad 
\sin\theta_{\omega}=\sqrt{\frac{1}{2}\left(1-\frac{P^z}{|\vec{P}|}\cos{2\theta}\right)}~. 
\end{equation} 
 
Now, let us assume for a moment that we have more $\nu_e$ than $\nu_x$ so that $P^z>0$. In this case, Eq. (\ref{Angles in the Infinity Limit}) yields $\cos{\theta_{\omega}}=\cos{\theta}$, i.e., the angle $\theta_{\omega}$ is equal to the vacuum mixing angle for all neutrinos. Also note that Eqs. (\ref{Transformation}) lead to $\mathcal{P}^+=\sin{2\theta}P^z$ which tells us that the phase $\delta$ defined in Eq. (\ref{delta}) is equal to zero for this particular case. These observations lead to the result that $z_{\omega}=z=\tan{\theta}$ for all neutrinos. In this case, the operator $U_{\bbomega}$ simply becomes equal to the operator ${U}$ defined in Eq. (\ref{Unitary Operator}). As a result, one arrives at  
\begin{equation} 
\label{Bogoliubov Transformed Particles in the Limit} 
\alpha_1(\mathbf{p})={U}^\dagger a_1(\mathbf{p}){U} =a_e(\mathbf{p})  
\qquad\mbox{and}\qquad  
\alpha_2(\mathbf{p})={U}^\dagger a_2(\mathbf{p}){U} =a_x(\mathbf{p})~, 
\end{equation} 
where we used Eq. (\ref{Unitary Transformation of Particle Operators}). The result stated in Eq. (\ref{Bogoliubov Transformed Particles in the Limit}) is important because it tells us that in the limit where $\mu\to\infty$, the  quasi-particle operators become equal to the flavor particle operators and thus the state in which all neutrinos occupy flavor states is an eigenstate of the RPA Hamiltonian (\ref{RPA Hamiltonian in b's}) in this limit. Note that, although we assumed that we have initially more $\nu_e$ than $\nu_x$, the opposite scenario can be worked out in a similar way and shown to lead to the same conclusion with $\alpha_1(\mathbf{p})=-a_x(\mathbf{p})$ and $\alpha_2(\mathbf{p})=a_e(\mathbf{p})$.

Since one of the states $|\psi\rangle$ in Eq. (\ref{RPA Eigenstates}) represents the initial state of the system, then we have a steady-state case with no time evolution as long as $\mu$ stays constant. It is already known that in the limit where $\mu\to\infty$ the state in which all neutrinos occupy flavor eigenstates is a steady-state solution which is consistent with the above discussion. However, if $\mu$ decreases (which happens when the neutrinos emanate from a source), then the quasi-particle operators $\alpha_1(\mathbf{p})$ and $\alpha_2(\mathbf{p})$ change accordingly and the initial state cannot remain as an eigenstate. However, if the change in $\mu$ occurs in an adiabatic way, then the state evolves steadily in such a way that it continues to keep its original form in terms of the quasi-particles while the quasi-particle operators $\alpha_1(\mathbf{p})$ and $\alpha_2(\mathbf{p})$ themselves slowly change. In the limit where $\mu\to 0$, Eq. (\ref{lambda_w}) yields 
\begin{equation} 
2\lambda_{\omega}=|\omega_c-\omega| 
\end{equation} 
and Eqs. (\ref{angles}) lead to  
\begin{equation} 
\cos\theta_{\omega}=\sqrt{\frac{1}{2}\left(1+\frac{\omega_c-\omega}{|\omega_c-\omega|}\right)} \qquad 
\sin\theta_{\omega}=\sqrt{\frac{1}{2}\left(1-\frac{\omega_c-\omega}{|\omega_c-\omega|}\right)}~. 
\end{equation} 
As a result, in $\mu\to 0$ limit one obtains $\cos{\theta_{\omega}}=1$ and $\sin{\theta_{\omega}}=0$ for those neutrinos with $\omega<\omega_c$ leading to 
\begin{equation} 
\alpha_1(\mathbf{p})=a_1(\mathbf{p})~, \qquad\mbox\qquad \alpha_2(\mathbf{p})=a_2(\mathbf{p})~. 
\end{equation} 
On the other hand for those neutrinos with $\omega>\omega_c$, the $\mu\to 0$ limit yields $\cos{\theta_{\omega}}=0$ and $\sin{\theta_{\omega}}=1$ leading to 
\begin{equation} 
\alpha_1(\mathbf{p})=a_2(\mathbf{p})~, \qquad\mbox\qquad \alpha_2(\mathbf{p})=a_1(\mathbf{p})~, 
\end{equation} 
up to some inconsequential phase factors. We see that, a state which starts off with all neutrinos in flavor eigenstates in the $\mu\to\infty$ limit can adiabatically evolve towards the $\mu\to 0$ limit in such a way that those neutrinos with $\omega<\omega_c$  and those with $\omega>\omega_c$ end up in the opposite mass eigenstates. For example, if we initially have more $\nu_e$ than $\nu_x$, then those electron neutrinos with $\omega<\omega_c$ evolve into the first mass eigenstate while those with $\omega>\omega_c$ evolve into the second mass eigenstate. In this case, the opposite is true for $\nu_x$. Such an evolution leads to a spectral split as discussed in Ref. \cite{Raffelt:2007cb}.  
 
In order to find the split frequency $\omega_c$, one should solve the consistency equations given in Eqs. (\ref{Consistency Conditions}).  Since we assumed that initially all neutrinos are in flavor states, one has $2m_{\omega}=P_{\omega}^z=\epsilon_{\omega}|\vec{\mathcal{P}}_{\omega}|$ where $\epsilon_{\omega}=\pm 1$ depending on whether $P_{\omega}^z$ initially points up (more $\nu_e$) or down (more $\nu_x$). As a result, the consistency equations given in Eq. (\ref{Consistency Conditions}) reduce to 
\begin{equation} 
\label{Consistency Conditions - Special} 
1=\sum_{\omega}\frac{\epsilon_{\omega}|\vec{\mathcal{P}}_{\omega}|}{\sqrt{(\frac{\omega_c-\omega}{\mu}+\mathcal{P}^z)^2+\mathcal{P}^+\mathcal{P}^-}} 
\qquad\mbox{and}\qquad 
\omega_c=\sum_{\omega} \frac{\epsilon_{\omega}\omega|\vec{\mathcal{P}}_{\omega}|}{\sqrt{(\frac{\omega_c-\omega}{\mu}+\mathcal{P}^z)^2+\mathcal{P}^+\mathcal{P}^-}}~. 
\end{equation} 
These equations were also obtained in Ref. \cite{Raffelt:2007cb} where a method based on the time evolution of polarization vectors is used. The derivation presented here in terms of the evolution of quasi-particles is physically equivalent to the one given in Ref. \cite{Raffelt:2007cb}.  
 
\section{Antineutrinos} 
\label{Section: Antineutrinos} 
 
In this section, we include antineutrinos in the formalism introduced in previous sections. Let us denote the fermion operators for a neutrino with momentum $\mathbf{p}$ in the flavor state $\bar{\nu}_\alpha$ by $\bar{a}_{\alpha}\left(\mathbf{p}\right)$ and in the mass state $\bar{\nu}_i$ by $\bar{a}_{i}\left(\mathbf{p}\right)$ where $\alpha=e,x$ and $i=1,2$. It has been pointed out by many authors that describing antineutrinos with the \emph{rotated} spinor $(-\bar{\nu}_2, \bar{\nu}_1)$ instead of the regular spinor $(\bar{\nu}_1, \bar{\nu}_2)$ is more advantageous (see, for example, Refs. \cite{Duan:2007fw, Duan:2007bt}) because 
this leads to a seamless integration of antineutrinos into the formalism making the physics more transparent (see Appendix \ref{Appendix Antineutrino Hamiltonian}). In order to implement this, one can first define the antineutrino isospin operators as in Eqs. (\ref{Flavor Isospin Operators})  and (\ref{Mass Isospin Operators}) and then perform a transformation as in Eqs. (\ref{General Unitary Transformation}) with $\theta_\mathbf{p}=\pi/2$ and $\delta_\mathbf{p}=0$ for all momentum modes\footnote{Note that for $\theta_\mathbf{p}=\pi/2$ the parameter $z_\mathbf{p}$ is singular in Eq. (\ref{General Unitary Operator}). However, the transformation is still well defined because the operator $U_\mathbb{p}$ can be written in several alternative forms. One of them, which is not singular for $\theta_\mathbf{p}=\pi/2$, is given by 
\begin{equation*} 
{U_{\mathbb{p}}}=e^{\sum_{\mathbf{p}}2\theta_\mathbf{p}\left(\mathcal{J}^x\sin\delta_\mathbf{p}+\mathcal{J}^y\cos\delta_\mathbf{p}\right)}~.
\end{equation*} 
}. This leads to 
\begin{equation}
\label{Rotated Particle Operators} 
\tilde{a}_1({\mathbf{p}}) \equiv -\bar{a}_2({\mathbf{p}})~, 
\qquad\qquad 
\tilde{a}_2({\mathbf{p}})\equiv \bar{a}_1({\mathbf{p}}) 
\end{equation}
and similarly for the particle operators of flavor states. We denote the corresponding flavor and mass antineutrino isospin operators by ${\vec{\tilde{J}}}_{\mathbf{p}}$ and ${\vec{\tilde{\mathcal{J}}}}_{\mathbf{p}}$ which are defined as in Eqs.  (\ref{Flavor Isospin Operators})  and (\ref{Mass Isospin Operators}), respectively.

With the introduction of antineutrinos, we modify the summation convention introduced in Eq. (\ref{Totals}) as 
\begin{equation} 
A_{\omega}\equiv\sum_{|\mathbf{p}|=p}A_{\mathbf{p}}, \qquad A_{-|\omega|}\equiv\sum_{|\mathbf{p}|=p}\tilde{A}_{\mathbf{p}} \qquad \mbox{and} \qquad 
A\equiv\sum_{\omega} A_{\omega}~ 
\label{Totals - Anti} 
\end{equation}
and also introduce 
\begin{equation}
\label{Define -w}
\omega=-\frac{\delta m^2}{2p}
\end{equation}
for antineutrinos. Note that the last sum in Eq. (\ref{Totals - Anti}) over $\omega$ now includes the negative values for antineutrinos as well as the positive values for neutrinos. 

The Hamiltonian describing the vacuum oscillations and self interactions of neutrinos and antineutrinos in the single angle approximation is given by Eq. (\ref{Hamiltonian}), except that now the new convention introduced in Eq. (\ref{Totals - Anti}) applies \cite{Sawyer:2005jk}. As a result, all of the main results of this paper, including the invariants given in Eqs. (\ref{Invariants}) and (\ref{Invariants RPA}), can be generalized to include antineutrinos by extending the sums over $\omega$ to include negative values. We only would like to make a few comments in what follows. 

With the inclusion of antineutrinos, the number of invariants is now doubled because one has the invariants $h_\omega$ for neutrinos and the invariants $h_{-\omega}$ for antineutrinos. Considering the diagonalization of the Hamiltonian described in Section (\ref{Subsection: The Bethe Ansatz Method}), $j$ now represents the total isospin quantum number of both neutrinos and antineutrinos in accordance with Eq. (\ref{Totals - Anti}). In the case of the electrostatic analogy, one has additional negative charges $-j_{-\omega}$ located on the negative $x$-axis at positions $-\omega$. In this case, one can still apply the formalism described in Section \ref{Subsection: Electrostatic Analogy} by shifting the origin of the coordinate system in Fig. \ref{Fig: Electrostatic_Analogy}. Finally we note that when antineutrinos are treated as described above, their polarization vectors defined as in Eq. (\ref{Polarization Vectors}) are given by  
\begin{equation} 
\label{Polarization Vectors Explicit - Anti} 
\vec{\tilde{P}}_{\mathbf{p}}= 
\left( 
\begin{array}{c} 
-2 \mbox{Re}[\bar{\psi}_x^*(\mathbf{p}) \bar{\psi}_e(\mathbf{p})]   \\ 
-2 \mbox{Im}[\bar{\psi}_x^*(\mathbf{p}) \bar{\psi}_e(\mathbf{p})]  \\ 
|\bar{\psi}_x(\mathbf{p})|^2-|\bar{\psi}_e(\mathbf{p})|^2   
\end{array} 
\right)~. 
\end{equation}  
This definition involves an overall minus sign in comparison to the one which is frequently used in the literature.

\section{Conclusions} 
\label{Section: Conclusions} 
 
In this paper, we studied the symmetries and the associated constants of motion of the collective neutrino oscillation Hamiltonian by taking  into account both the vacuum oscillations and the self interactions of neutrinos. We examined the system both from the exact many-body perspective and from the point of view of an effective one-body description formulated with the application of the RPA method. We showed that, under the single angle approximation, both the many-body and the RPA pictures possess many constants of motion manifesting the existence of associated dynamical symmetries in the system.  
 
The existence of these constants of motion make the system completely integrable in the sense that the exact eigenstates and eigenvalues can be found in an analytical way. We wrote down these eigenstates and eigenvalues for both the exact many-body and the RPA Hamiltonians.  In the case of the many-body Hamiltonian, the eigenvalues and eigenstates were found with the application of the Bethe ansatz method and they depend on the solutions of Bethe ansatz equations which are analogous to the equilibrium conditions of an electrostatic system in two dimensions. In the case of the RPA Hamiltonian, the eigenstates and eigenvalues  were found by applying a suitable Bogoliubov transformation which brings the Hamiltonian into the form of a noninteracting system in terms of quasi-particles. We wrote down the consistency equations for all RPA eigenstates and showed that these equations reduce to a particular but physically very important case which was studied earlier in Ref. \cite{Raffelt:2007cb} in connection with the spectral splits of neutrinos. We showed that the spectral splits can be understood as the adiabatic evolution of quasi-particle states from a high density region where they coincide with flavor eigenstates into the vacuum where they coincide with mass eigenstates. 
 
In general, the existence of constants of motion offers practical ways of extracting information even from exceedingly complex systems. Even when the symmetries which guarantee their existence is broken, they usually provide a convenient set of variables which behave in a relatively simple manner depending on how drastic the symmetry breaking factor is. In this paper, we omitted an ordinary matter background and concentrated on a neutrino gas which undergoes vacuum oscillations as well as self interactions. Although an ordinary background of $p$, $n$, $e^-$ and $e^+$ coexists with neutrinos in most astrophysical sites, it is only the net electron fraction which plays a part in flavor evolution of neutrinos through a diagonal potential in flavor basis which is proportional to $\sqrt{2}G_F|n_{e^-}-n_{e^+}|$. Neglecting this potential can only be justified when the asymmetry of the electron background is small in comparison to the neutrino background. On the other hand, even under the circumstances where the electron background asymmetry cannot be ignored, the constants of motion presented in this paper can still be useful as a set of convenient variables because, since they commute with the \emph{neutrino gas} part of the Hamiltonian, their time variation only come from their commutator with the electron background term.  
 
As was recently illustrated in Ref. \cite{Raffelt:2011yb}, existence of such invariants naturally leads to associated $N$-mode collective neutrino oscillations. These collective oscillations are closely related to the $m$-spin solutions presented in Ref. \cite{yuzb} in the context of the BCS model. However, symmetries alone do not guarantee the stability of such collective behavior. Some issues related to the stability of the $m$-spin solutions of the BCS model have been addressed in Refs. \cite{yuzb,yuzb2}. In the case of the collective neutrino phenomena associated with our invariants the question of stability is an open problem and could be illuminated by numerical studies.

\section*{Acknowledgments} 
 
A.B.B. and Y.P. wish to thank the National Astronomical Observatory of Japan for their hospitality while much of this work was being done.   
This work was supported in part 
by the U.S. National Science Foundation Grant No. PHY-0855082,  
in part by the University of Wisconsin Research Committee with funds 
granted by the Wisconsin Alumni Research Foundation, in part  
through JUSTIPEN (Japan-U.S. Theory Institute for Physics with Exotic Nuclei) under  
grant number DEFG02- 06ER41407 (U. Tennessee),  
in part by Grants-in-Aid for Scientific Research (20244035 and 20540284) and on Innovative Areas (20105004) of the Ministry of Education, Culture, Sports, Science and Technology of Japan, and also by JSPS Core-to-Core Program, International Research Network for Exotic Femto Systems (EFES). 
We are grateful to the referee for his insightful comments. 

\appendix 
 
\section{Alternative Ways of Expressing Quantum Invariants} 
\label{Appendix Invariants} 
 
In this appendix, we will briefly mention two alternative ways to rewrite the constants of motion mentioned in the text. The first method is based on the following sum of the many-body invariants ${h}_{\omega}$ and ${L}_{\omega}$: 
\begin{equation} 
\label{Sum of Invariants 1} 
{C}_n =\sum_{\omega}\omega^n{h}_{\omega}+ 
\sum_{\omega} n\omega^{n-1}{L}_{\omega}~. 
\end{equation} 
Clearly ${C}_n$ is a constant of motion for every value of $n$ but here we specifically assume that $n$ is a positive integer. If we define the quantities  
\begin{equation} 
\vec{Q}_n = \sum_{\omega} \omega^n \vec{J}_{\omega}~, 
\end{equation} 
then ${C}_n$ defined in Eq. (\ref{Sum of Invariants 1}) can be written as 
\begin{eqnarray} 
\label{Invariant Alternative 1} 
{C}_n = \vec{B}\cdot\vec{Q}_n+ \mu\left(\vec{Q}_{n-1} \cdot \vec{Q}_{0} + \vec{Q}_{n-2} \cdot \vec{Q}_{1} + \dots + \vec{Q}_{1} \cdot \vec{Q}_{n-2} + \vec{Q}_{0} \cdot \vec{Q}_{n-1}\right)~. 
\end{eqnarray} 
The invariant ${C}_0$ was already mentioned in Eq. (\ref{Q^z Combination}). The Hamiltonian itself is equal to ${C}_1$ as can be seen from Eq. (\ref{C1}). 
 
A similar method can be used to rewrite the constants of motion of the RPA formalism. One starts from the sum of the RPA invariants $I_{\omega}$ and $|\vec{P}_{\omega}|^2$ given by 
\begin{equation} 
\label{Sum of Invariants 1 RPA} 
K_n=\sum_{\omega}\omega^n I_{\omega}+
\frac{1}{2}\sum_{\omega} n\omega^{n-1}|\vec{P}_{\omega}|^2
\end{equation} 
which is clearly an invariant for every $n$. Defining the quantities 
\begin{equation} 
\vec{P}_n = \sum_{\omega} \omega^n \vec{P}_{\omega}~, 
\end{equation} 
it can be shown that $K_n$ is equal to 
\begin{equation} 
\label{Invariant Alternative 1 RPA} 
K_n = \vec{B}\cdot\vec{P}_n+ \frac{\mu}{2}\left(\vec{P}_{n-1} \cdot \vec{P}_{0} + \vec{P}_{n-2} \cdot \vec{P}_{1} + \dots + \vec{P}_{1} \cdot \vec{P}_{n-2} + \vec{P}_{0} \cdot \vec{P}_{n-1}\right)~. 
\end{equation} 
 
Another alternative way to express the many-body quantum invariants, which is frequently encountered in the literature on Gaudin formalism, is based on the following sum: 
\begin{equation} 
\label{Sum of Invariants 2} 
{H}(\lambda)=-\sum_{\omega}\frac{{h}_{\omega}}{\omega-\lambda}+\sum_{\omega}\frac{{L}_{\omega}}{(\omega-\lambda)^2}~. 
\end{equation} 
Here $\lambda$ is a complex valued parameter and ${H}(\lambda)$ is clearly a constant of motion for every value of $\lambda$. Defining 
\begin{equation} 
\vec{\mathcal{Q}}(\lambda)=\sum_{\omega} \frac{\vec{\mathcal{J}}_{\omega}}{\omega-\lambda}
\end{equation} 
and using Eq. (\ref{Invariants}), one can show that the operator ${H}(\lambda)$ defined in Eq. (\ref{Sum of Invariants 2}) can be written as 
\begin{equation} 
\label{Invariant Alternative 2}  
{H}(\lambda)=\mathcal{Q}^z(\lambda)+\mu\vec{\mathcal{Q}}(\lambda)\cdot\vec{\mathcal{Q}}(\lambda)~. 
\end{equation} 
${H}(\lambda)$ is said to form a \emph{one parameter family of conserved quantities}. It satisfies 
\begin{equation} 
\left[ {H}(\lambda), {H}(\mu)\right] =0 \qquad \mbox{and} \qquad \left[ {H}(\lambda), {H} \right] =0  
\end{equation}  
for every complex value of $\lambda$ and $\mu$. 
 
In the case of the RPA formalism, one similarly defines the sum of the RPA invariants $I_{\omega}$ and $|\vec{P}_{\omega}|^2$ as follows: 
\begin{equation} 
D(\lambda)=-\sum_{\omega}\frac{I_{\omega}}{\omega-\lambda}+\frac{1}{2}\sum_{\omega}\frac{|\vec{P}_{\omega}|^2}{(\omega-\lambda)^2}~. 
\end{equation} 
$D(\lambda)$ is a constant of motion for every complex value of $\lambda$. Defining the vectors 
\begin{equation} 
\vec{\mathcal{P}}(\lambda)=\sum_{\omega} \frac{\vec{\mathcal{P}}_{\omega}}{\omega-\lambda}~, 
\end{equation} 
one can show that it can be written as 
\begin{equation} 
\label{Invariant Alternative 2 RPA}  
D(\lambda)=\mathcal{P}^z(\lambda)+\frac{\mu}{2}\vec{\mathcal{P}}(\lambda)\cdot\vec{\mathcal{P}}(\lambda)~. 
\end{equation} 
$D(\lambda)$ also forms a one parameter family of conserved quantities.  

\section{Antineutrino Hamiltonian} 
\label{Appendix Antineutrino Hamiltonian} 

The effective many-body Hamiltonian describing the collective flavor evolution of neutrinos in a dense medium in the presence of neutrino self interactions was obtained in Ref. \cite{Sawyer:2005jk} by expanding the full neutral current interaction and keeping only the forward scattering terms. The underlying group is $SU(3)\times SU(3)$ representing all three neutrino and antineutrino flavors. Reducing to the $SU(2)\times SU(2)$ subgroup for the two neutrino case yields the following effective many-body Hamiltonian:
\begin{eqnarray}
H&=&\sum_{\omega>0}\omega\left(-\cos2\theta\: J_{\omega}^{0}+\sin2\theta\:\frac{J_{\omega}^{+}+J_{\omega}^{-}}{2} -\cos2\theta\:\bar{J}_{-\omega}^{0}+\sin2\theta\:\frac{\bar{J}_{-\omega}^{+}+\bar{J}_{-\omega}^{-}}{2}\right)
\label{Total Hamiltonian Unrotated in Flavor Basis}
\\ & + & 
\mu
\sum_{\mathbf{p},\mathbf{q}}(1-\cos\vartheta_{\mathbf{p}\mathbf{q}})\left(\vec{J}_{\mathbf{p}}\cdot\vec{J}_{\mathbf{q}}
+
\vec{\bar{J}}_{\mathbf{p}}\cdot\vec{\bar{J}}_{\mathbf{q}}
-
2J_{\mathbf{p}}^{0}\bar{J}_{\mathbf{q}}^{0}-J_{\mathbf{p}}^{+}\bar{J}_{\mathbf{q}}^{+}-J_{\mathbf{p}}^{-}\bar{J}_{\mathbf{q}}^{-}\right)
~. \nonumber
\end{eqnarray}
Here $\vec{\bar{J}}$ are the flavor isospin operators for antineutrinos defined as in Eq. (\ref{Flavor Isospin Operators}). They are subject to the same summation convention as $\tilde{A}$ given in Eq. (\ref{Totals - Anti}).  
The terms involving the neutrinos and antineutrinos in the Hamiltonian (\ref{Total Hamiltonian Unrotated in Flavor Basis}) are dissimilar. However, if one performs the transformation described at the beginning of Section \ref{Section: Antineutrinos} and defines the antiparticle operators $\tilde{a}$ as in Eq. (\ref{Rotated Particle Operators}), then the corresponding isospin operators are related by
\begin{equation}
(\tilde{J}^0_{\mathbf{p}},\; \tilde{J}^+_{\mathbf{p}}, \;\tilde{J}^-_{\mathbf{p}})=
(-\bar{J}^0_{\mathbf{p}},\; -\bar{J}^-_{\mathbf{p}}, \; -\bar{J}^+_{\mathbf{p}})~.
\end{equation}
In terms of the transformed antineutrino isospin operators, the Hamiltonian in Eq. (\ref{Total Hamiltonian Unrotated in Flavor Basis}) becomes 
\begin{eqnarray}
H&=&\sum_{\omega>0}\omega\left(-\cos2\theta\: J_{\omega}^{0}+\sin2\theta\:\frac{J_{\omega}^{+}+J_{\omega}^{-}}{2} +\cos2\theta\:\tilde{J}_{-\omega}^{0}-\sin2\theta\:\frac{\tilde{J}_{-\omega}^{+}+\tilde{J}_{-\omega}^{-}}{2}\right)
\label{Total Hamiltonian Rotated in Flavor Basis}
\\ & + & 
\mu
\sum_{\mathbf{p},\mathbf{q}}(1-\cos\vartheta_{\mathbf{p}\mathbf{q}})\left(\vec{J}_{\mathbf{p}}\cdot\vec{J}_{\mathbf{q}}
+
\vec{\tilde{J}}_{\mathbf{p}}\cdot\vec{\tilde{J}}_{\mathbf{q}}
+
2\vec{J}_{\mathbf{p}}\cdot\vec{\tilde{J}}_{\mathbf{q}}\right)
~. \nonumber
\end{eqnarray}
If we define the oscillation frequencies of antineutrinos with a minus sign as given in Eq. (\ref{Define -w}), then (in the single angle approximation) the total Hamiltonian (\ref{Total Hamiltonian Rotated in Flavor Basis}) takes the form of Eq. (\ref{Hamiltonian}) with the sum over $\omega$ extended to include the negative values.  


\end{document}